\documentclass[12pt,preprint]{aastex}
\usepackage{amsmath}

\begin{document}

\title{Rotating Motions and Modeling of the Erupting Solar Polar Crown Prominence on 2010 December 6}

\author{Yingna Su\altaffilmark{1}, Adriaan van Ballegooijen\altaffilmark{1}}
\altaffiltext{1}{Harvard-Smithsonian Center for Astrophysics, Cambridge, MA 02138, USA.}  
\email{ynsu@head.cfa.harvard.edu}

\begin{abstract}

A large polar-crown prominence composed of different segments spanning nearly the entire solar disk erupted on 2010 December 6. 
Prior to the eruption, the filament in the active region part splits into two layers: a lower layer and an elevated layer. 
The eruption occurs in several episodes. Around 14:12 UT, the lower layer of the active region filament  breaks apart, one part ejects towards the west, while the other part ejects towards the east, which leads to the explosive eruption of the eastern quiescent filament. During the early rise phase, part of the quiescent filament sheet displays strong rolling motion (observed by STEREO$\_$B) in the clockwise direction (views from east to west) around the filament axis. This rolling motion appears to start from the border of the active region, then propagates towards the east. AIA observes another type of rotating motion: in some other parts of the erupting quiescent prominence the vertical threads turn horizontal, then turn upside down. The elevated active region filament does not erupt until 18:00 UT, when the erupting quiescent filament already reaches a very large height. We develop two simplified three-dimensional models which qualitatively reproduce the observed rolling and rotating motions.
The prominence in the models is assumed to consist of a collection of discrete blobs that are tied to particular field lines of a helical flux rope. The observed rolling motion is reproduced by continuous twist injection into the flux rope in Model 1 from the active region side. Asymmetric reconnection induced by the asymmetric distribution of the magnetic fields on the two sides of the filament may cause the observed rolling motion. The rotating motion of the prominence threads observed by AIA is consistent with the removal of the field line dips in Model 2 from the top down during the eruption.

\end{abstract}

\clearpage


\section{INTRODUCTION}

 Solar flares, filament eruptions and coronal mass ejections (CMEs) are spectacular solar events and are the primary driver of ``space weather"
 at Earth. It is well accepted that these three phenomena are different manifestations of a single physical process that involves a disruption of the  coronal magnetic field \citep{1996SoPh..166..441H, 2000JGR...10523153F}, but it is unclear how the eruption starts. Filaments or prominences are chromospheric material suspended in the corona by magnetic fields. We will use filament and prominence interchangebly throughout the text. Since the coronal magnetic field is difficult to observe, filaments can be useful tracers of the locations and movements of the erupting fields \citep{2005ApJ...630.1148S}. Therefore, studying the evolution of the erupting filament may shed light on the trigger mechanism of a specific eruption.
  
Helical structures are frequently observed  during prominence eruptions \citep[][ and references therein]
{1988SoPh..116...45V, 1994SoPh..155...69R, 2012A&A...540A.127K}. 
Helical-like structures are also observed in filaments before the eruption \citep[e.g., ][]{1987SoPh..108..251K}
or in stable filaments \citep[e.g., ][]{1975STIN...7614007R, 1993SoPh..146..147V}.
Sometimes helical structures are visible only in part of the
prominence (internal/microscopic magnetic twist). In other cases, especially during eruptions,
they are the most visible feature (external/macroscopic twist), such as in ``helical prominences'' where the prominence body is
composed of a bundle of twisted threads \citep[e.g., ][]{1975STIN...7614007R, 1988SoPh..116...45V, 
1991SoPh..136..151V, 1991SoPh..133..339S}.  

The frequent exhibition of helical structures in erupting filaments is evidence of 
a twisted magnetic flux rope. There are two types of models regarding the formation of the twisted flux rope.  
Some models assume that a twisted flux rope is present in the region before the eruption. 
This flux rope becomes unstable as a result of footpoint motion, injection of magnetic helicity, and/or draining of heavy
prominence material \citep{1991ApJ...373..294F, 1998ApJ...493..460G, 2000ApJ...539..964K, 1997SoPh..170..265W,
2003ApJ...588L..45R}. Other models begin with an untwisted, but highly sheared
magnetic field that becomes unstable as a result of reconnection, flux
cancellation, or similar process \citep{1988ApJ...328..830M, 1994ApJ...430..898M, 1999ApJ...510..485A,
2003ApJ...585.1073A, 2003JGRA..108.1162M}. According to the latter
models, a twisted flux rope does not exist prior to eruption, but is
formed during the eruption as a result of reconnection between
the two sides of the sheared arcade. Therefore, in both types of
models a twisted flux rope is present in the ejecta. The existence of
such flux ropes is confirmed by in situ observations of magnetic
fields in interplanetary magnetic clouds \citep{1991pihp.book....1B}. However,
the question remains when and how such flux ropes are formed.

Erupting filaments are sometimes observed to undergo a rotation about the vertical direction as they rise
\citep[e.g., ][]{1987SoPh..108..251K, 2003ApJ...595L.135J, 2006ApJ...651.1238Z,  2007SoPh..246..365G, 2007ApJ...661.1260L, 2009ApJ...697..999L,2009ApJ...703..976M, 2011A&A...531A.147B, 2011ApJ...730..104J, 2011JASTP..73.1138T}. 
This filament rotation is interpreted as a conversion of twist into writhe in a kink-unstable flux rope. Therefore, MHD
helical-kink instability is often taken to be the primary trigger of these eruptions \citep{2007ApJ...668.1232F, 2010A&A...516A..49T, 2012SoPh..tmp...91K}. Consistent with this interpretation, the rotation is usually found to be clockwise (as viewed from above) if the post-eruption arcade has right-handed helicity, but
counterclockwise if it has left-handed helicity. Helicity is a quantitative, mathematical measure of chirality 
\citep{1984JFM...147..133B, 2009ApJ...703..976M, 2011A&A...531A.147B}. Magnetic reconnection
with ambient field can also cause filament rotation during eruption \citep{2010JGRA..11510104C, 2011JASTP..73.1138T}.  
The aforementioned filament rotation should be distinguished from the rotation of the filament around its own axis, namely the ``roll effect"
\citep{2003AdSpR..32.1883M, 2011JASTP..73.1129P}, i.e., the top of the prominence spine gradually bends to one side of the spine during the rise of
the filament. This sideways rolling creates twist of opposite sign in the two prominence legs as the prominence continues to rise.
\citet{2011JASTP..73.1129P} interpreted this effect as consequence of force imbalances inside the filament arcade related to the adjacent large
coronal holes. \citet{2012ApJ...751...56M} hypothesize that this rolling motion is induced by an offset between the CME current sheet and the rising
flux rope during some events.  

In the present work, we identify two other types of filament rotation that occurred during the filament eruption on 2010 December 6.
\citet{Thompson2012} presents STEREO/EUVI observations of this filament, and divides this filament into three branches. The main eruption occurred in the middle branch around 14:16 UT on December 6, while the right part of the filament erupted at 2:06 UT on December 7.  \citet{Thompson2012} studied the magnetic twist in the erupting prominence using triangulation of prominence threads with primary focus on the second eruption. The structure and dynamics of this prominence before the eruptions are presented in our previous paper \citep[][hereafter we will call it Paper~I]{2012ApJ...757..168S}. In the current paper we focus on plasma dynamics during the first eruption, which will be 
referred as filament eruption or main eruption on December 6 in the following sections.

\section{Observations}

A polar-crown filament spanning nearly the entire solar disk erupted around 14:16 UT on 2010 December 6. This filament eruption is associated with a CME, and a corresponding long-duration B2 (GOES class) flare is identified. The SOHO LASCO CME catalog\footnote{http://cdaw.gsfc.nasa.gov/CME$\_$list/} shows that the CME first appears in the LASCO/C2 field of view at 17:24 UT, and the linear speed is 538 km s$^{-1}$ projected onto the plane of sky. According to the CACTus CME catalog\footnote{http://sidc.oma.be/cactus/catalog.php}, the CME first appears at LASCO/C2 at 18:00 UT, and the CME median velocity is 523 km s$^{-1}$.

The prominence eruption under study is well observed near the east limb by the Atmospheric Imaging Assembly \citep[AIA, ][]{2012SoPh..275...17L} aboard the $\emph{Solar Dynamics Observatory}$. EUVI (Extreme Ultraviolet Imaging Telescopes) onboard STEREO (Solar Terrestrial Relations Observatories)\citep{2004SPIE.5171..111W, 2008SSRv..136...67H} also observed this eruption. The separation angle between the twin STEREO spacecraft was about 171$^{\circ}$. STEREO$\_$B (Behind) viewed this event on the solar disk, while this event is visible at the east limb in the view of STEREO$\_$A (Ahead). Synoptic observations by the X-Ray Telescope \citep[XRT, ][]{2007SoPh..243...63G, 2008SoPh..249..263K} aboard $\emph{Hinode}$ \citep{2007SoPh..243....3K} and H$\alpha$ observations by KSO (Kanzelh\"ohe Solar Observatory) are also included. The photospheric magnetic field information is provided by the Helioseismic and Magnetic Imager \citep[HMI, ][]{2012SoPh..275..229S} aboard SDO.  

The large polar-crown filament is composed of active region parts and quiet Sun parts on the eastern and western sides of the active region.
AIA and STEREO$\_$B images of the prominence at 304~\AA~prior to the eruption are shown in Figure 1. According to the eruption behavior, 
the prominence is divided into 7 different segments, which are marked using white arrows in Figure 1. Prior to the eruption, the filament in the active region split into two layers.
One layer lies at the core of the active region (AR1), while the other layer is elevated (AR2). This type of double-decker filament has also been studied by \citet{2012ApJ...756...59L}. The quiescent prominence on the eastern side of the active region is composed of 4 segments. The segment closest to the active region is named Q1, and the other segments are named Q2, Q3, and Q4 sequentially based on the distance away from the active region (See Figure 1). Q2 and Q3 prominences appear to be divided by the dense column (DC) structure (black arrow in Figure 1), which is similar to the solar ``tornado" studied by \citet{2012ApJ...756L..41S}. The non-erupting segment Q4 and the quiescent filament on the western side of the active region (Q5) are corresponding to the left and right filament branches as named by \citet{Thompson2012}. \citet{Thompson2012} presented that the eruption occurs in the middle branch of the filament, which is corresponding to the AR1, AR2, Q1, Q2, and Q3 filaments in the current paper.

Figures 2--4 show AIA, STEREO$\_$B/EUVI, and  STEREO$\_$A/EUVI observations of the prominence at the onset of (top row), during (middle row), and after (bottom row) the eruption at 304~\AA~(left column) and 193/195~\AA~(right column) on December 6. The eruption originates from brightenings and ejections of AR1 filament material in two opposite directions (east and west). This ejection leads to the eruption of the eastern quiescent filament (Q1, Q2, Q3) on December 6 (Figures 2c, 3c, 4c), which is followed by a slow rise of the western quiescent filament (Q5) that erupted on December 7 (Figures 2e, 3e, 4e). In the present study, we focus on the eruption on December 6. The nearest eastern quiescent prominence (Q1) erupts firstly and appears to be the strongest. The further the prominence is away from the active region, the later and weaker the eruption is. At the early stage of the eruption, Q1 and Q2 prominence displays a rolling motion while rising up as observed at 195~\AA~by STEREO$\_$B. Q1 prominence erupts like a jet or surge with a more inclined path rather than a radial path. It appears to be the first part to leave the AIA field of view.  Prominence Q2 evolves into an arch-like structure (Figures 2c, 3c, 4c) before its successful eruption. AIA has a better view of the Q3 prominence, which rises up then drains back to Q4 prominence which shows no eruption at all (Figures 2c, 2e). The arch-like structure (Q2) remains separate from the vertical threads (Q3) during the entire eruption process (Figure 2c). The overlying active region filament (AR2) rises up and erupts when the eastern quiescent filament has already risen to a much larger height. A timeline of various activities during the eruption is shown in Table 1.

Corresponding to Figures 2--4, three videos (videos 1-- 3) of this eruption are also available online in the electronic edition of the Journal. The cadence of the AIA video is 2 minutes which is reduced from the original 12-second observations. The cadence of STEREO observations at 195~\AA~is generally 5 minutes with an exception of 2-3 minutes between 16:10 UT and 20:00 UT for STEREO$\_$A. STEREO generally took images every 10 minutes at 304~\AA~with an exception of 2-3 minutes between 17:16 UT and 21:56 UT for STEREO$\_$A. Structure and dynamics of different prominence segments during the eruption are better viewed in the online videos. In the following sections, we will present detailed observations of the dynamics of different prominence segments during the eruption.

\subsection{Dynamics of the Active Region Filament (AR1 and AR2)}

STEREO$\_$A and AIA observations of the active region filament at the early stage of the eruption are shown in Figure 5. The eruption begins with the appearance of brightenings (white arrows in Figure 5d) on the two sides of the active region filament around 14:16 UT as observed at 193~\AA~by AIA. With the extension of the brightenings along the direction parallel to the filament (see video 1), the core active region filament (AR1) begins to rise up (Figure 5a) which leads to a slight rise of the elevated filament (AR2) (see videos 2 and 3). The AR1 filament seen at 304~\AA~by STEREO$\_$A starts to eject horizontally in two directions (marked with black arrows in Figure 5) after reaching the overlying filament (AR2) around 14:46 UT (top row in Figure 5). The major part ejects towards the east (Q1) which leads to the explosive eruption of the quiescent prominence on the eastern side of the active region which is the primary focus of the present study.  A small portion of the filament ejects towards the west (Q5) followed by an eruption at 2:06 UT on December 7, which is studied in detail by \citet{Thompson2012}.

Although the eruption originates from the active region part, the elevated filament in the active region does not erupt until 18:00 UT or so. The eruption of the elevated active region filament is associated with the expansion or disappearance of the overlying loops as observed at 195~\AA~by STEREO (white arrows in the right columns of Figures 3--4). Following the eruption of the active region filament, bright ribbons appeared on the two sides of the filament channel around 18:30 UT as observed at 304~\AA~by AIA (Figure 2e). Highly sheared post-eruption loops (Figure 2f) firstly appear at 193~\AA~by AIA around 19:30 UT. These loops show that the filament channel has sinistral orientation of the axial field (see Paper~I).

\subsection{Rolling Motion of Prominence Q1 and Q2 Observed by STEREO$\_$B}

After 14:21 UT, STEREO$\_$B/EUVI detects a clear rise of the elevated filament in the active region part at 195~\AA. Following this rise, the dark quiescent filament (Q1 and Q2) sheet adjacent to the active region begins to rise up as well and displays a rotating motion around the filament axis. This rolling motion is along the clockwise direction when viewing the filament from the east to the west (see video 2). Note that this rolling motion is well observed by STEREO$\_$B, while no clear corresponding motion is identified in the SDO/AIA and STEREO$\_$A observations. A series of STEREO$\_$B images at 195~\AA~of the early stages of the rolling motion are shown in Figure 6. The bottom and top of the filament sheet are marked using the solid and dashed white lines, respectively. At 195~\AA, the filament is observed in optically thick absorption, which makes it easier for us to qualitatively identify the relative heights of different layers. This figure shows that the western end of the bottom of the Q1 filament begins to rise upward, passing in front of the top of the sheet, which rotates downward gradually. Figure 6c shows that the western Q1 filament is turned around, i.e, the bottom of the sheet becomes the top and vice versa. This rolling motion (illustrated with black arrow in Figure 6) is associated with a slow rise of the filament. Around 14:51 UT, the projection of the filament sheet reaches the limb as shown by Figure 6d. STEREO$\_$B observations (video 2) show that this rolling motion begins from the border of the active region and quiet Sun, then propagates towards the quiescent filament on the east. This clear rolling motion appears mainly in the Q1 and Q2 filament segments. It lasts until 17:00 UT or so, after which the rising motion of the filament becomes dominant.

Initially, the bright curved features \citep[black arrow in Figure 3b, ][]{2010ApJ...721..901S, 2012ASPC..454..113S} on the northern side of the quiescent filament channel remain undisturbed. These bright features are interpreted as the lower legs of the field lines that turn into the flux rope in Paper~I. The bright features fades out starting from the eastern end (near the dense column structure), then spreads to the western part gradually. This process starts around 16:00 UT, and most of the bright curved features fade out by 19:00 UT.

\subsection{Rotating Motion of Q3 Prominence Observed by SDO}

The previous section shows the rolling motion of Q1 and Q2 filament segments as observed by STEREO$\_$B. Q3 prominence displays another type of rotating motion, which is clearly identified in the SDO observations at 304~\AA~as shown in Figure 7. The curved black arrows in Figure 7c represent the path of the prominence motion. This figure shows that Q3 contains mainly thin threads, which are initially crinkled and more vertical (Figure 7a), then turn horizontal and become more straight with time (Figure 7d). There is a lower edge for these threads. The southern end of this lower edge rises much faster than the northern end. Eventually, these initially vertical threads make a half turn while draining back to the Q4 filament (see Figure 7 and videos 1 and 2).  The northern leg of the arch of Q2 filament also shows the same rotating behavior, though not as clearly as that shown in Q3 filament.

Not all of the filament material successfully escaped from the Sun even for Q1 filament. Most Q3 filament material appears to fall back to the Sun as observed at 304~\AA~by AIA. The northern leg of Q2 filament displays very similar behavior as Q3 filament, though Q2 filament rises much higher. The falling motion starts from Q2 filament and is followed by the Q3 filament. Around 17:30 UT, a portion of materials in Q1 filament starts to fall back towards the Sun. Around 18:20 UT, the southern leg the Q2 filament arch shows sign of material falling which is associated with an untwisting motion, since all the falling material became straight in the end.  

\subsection{Post-eruption Arcades after the Eruption}

After the filament eruption on December 6, cusp-shaped arcades are observed by both SDO/AIA and Hinode/XRT as shown in Figure 8. The top three rows present AIA images at 211~\AA~(first row), 335~\AA~(second row), and 94~\AA~(third row) taken around 01:00 UT (left column) and 06:00 UT (right column) on 2010 December 7. In AIA, the cusp-shaped arcades are best observed at 94~\AA~(Figures 8e--8f), and they firstly appear around 20:00 UT on December 6. Cusp-shaped structures are also visible at 335~\AA~(Figures 8c--8d) after 1:00 UT on December 7. More rounded loop-like structures are observed in most of the other EUV channels (e.g., Figures 8a--8b). AIA observations suggest that the cusp-shaped arcades show clear evidence of field line shrinkage as that found by \citet{2008ApJ...675..868R}. The bottom row of Figure 8 shows two XRT images taken with Al-mesh and Ti-poly filters at 06:03 UT on December 7, respectively. This figure suggests that the cusp-shaped arcades are best viewed in the soft X-ray images taken by XRT. Moreover, XRT can see the more outer part of the cusp, i.e. the part that just goes through reconnection.

As mentioned earlier, the post-eruption loops firstly appear around 19:30 UT on December 6. Cusp structure is still visible 10 hours later (as shown in Figure 8), which indicates a slow and gradual reconnection event. The post-eruption arcades (Figure 8) appear to be a characteristic skewed candle flame shape. The skewed arcade is a signature of asymmetric reconnection as suggested by \citet{2012ApJ...751...56M}. The post-eruption arcades appear to be located at the quiet Sun region where the eruption of prominence Q1 and Q2 occurs. No clear post-eruption arcades are observed in the active region part. Moreover, a comparison of Figure 2 and Figure 8 shows that the straight features (i.e, lower legs of the large scale overlying loops, see black arrows in Figure 8) located on the two sides of the active region filament remain unchanged through the eruption. AIA and STEREO observations also show that the active region filament escapes towards the quiet Sun part on the east then erupts. These observations suggest that this eruption is a sideways eruption, because initially the filament did not erupt radially by opening up the overlying loops in the active region. A possible interpretation of the sideways or non-radial eruption is that the confining magnetic pressure decreases much faster horizontally than upward \citep{2012ApJ...757..149S}. 

\section{Model for the Erupting Prominence}

In this section two simple models for the dynamics of the erupting
prominence on December 6 are presented. The main goal is to understand the nature of
the rotating motions described in Section 2. 
The prominence is assumed to consist of a collection of discrete
blobs that are tied to particular field lines. The time-varying
magnetic field is described by a simple analytical model of a helical
flux rope, and the motion of the blobs is determined by solving their
equations of motion along the field lines, taking into account the
effects of gravity and friction with the surrounding corona. At the
top of the erupting flux rope the plasma is forced to move radially
outward, but in the legs the plasma falls back down to the
photosphere.  

\subsection{Assumptions and Methodology}

The prominence is assumed to be driven outward by an expanding
magnetic flux rope \citep[e.g.,][]{1997ApJ...490L.191C}. We assume that the flux
rope was already present prior to the eruption, and that the two legs
of the expanding flux rope remain attached to the photosphere. The
magnetic field is assumed to evolve according to ideal MHD, i.e.,
we neglect the effects of magnetic reconnection in the wake of the
erupting flux rope. However, we do take into account the writhing of
the flux tube \citep[e.g.,][]{2012SoPh..tmp...91K}. The shapes of the magnetic
field lines are described by an analytic function ${\bf r}
(p,q,\xi,t)$, where $p$ and $q$ are labels of the field lines,
$\xi$ is the coordinate along a field line, and $t$ is the time.
The coordinate $\xi$ varies from $-1$ at one endpoint of the field
lines to $+1$ at the other end. A detailed description of this
function and its free parameters is given in the Appendix.

The prominence plasma is described as a collection of discrete blobs
($i = 1, 2, \cdots$) located on different field lines characterized by
labels $p_i$ and $q_i$. The blobs are treated as point masses that are
tied to the particular field lines on which they are initially
located, i.e., a blob can slide along its field line, but cannot move
from one field line to another (consistent with our assumption of
ideal MHD). We simulate the dynamics of the blobs by solving their
equations of motion along the field lines, including the effects of
field-line motion, gravity, and a hypothetical frictional force. The
magnetic field is assumed to be relatively strong, so that the weight
of the prominence blobs does not significantly affect the shapes of
the field lines. In the initial state just before the start of the
eruption, the flux rope is assumed to be at rest in the low corona,
and the blobs are located at the dips of the helical windings.  The
expansion of the flux rope causes some blobs to be carried outwards by
the expanding field, while others fall back down to the
photosphere. 

Let $\xi_i(t)$ be the dimensionless coordinate of a blob along its
field line, then its position in the corona is ${\bf r}_i (t) =
{\bf r} [p_i,q_i,\xi_i(t),t]$. Each blob is carried along by the
motion of the field line on which it is located, so its velocity is
\begin{equation}
{\bf v}_i (t) \equiv \dot{\bf r}_i = \frac{\partial {\bf r}}
{\partial \xi} \dot{\xi}_i + \frac{\partial {\bf r}} {\partial t} ,
\end{equation}
and its acceleration is
\begin{equation}
\ddot{\bf r}_i = \frac{\partial {\bf r}} {\partial \xi} \ddot{\xi}_i +
\frac{\partial^2 {\bf r}} {\partial \xi^2} ( \dot{\xi}_i )^2 + 
2 \frac{\partial^2 {\bf r}} {\partial \xi \partial t} \dot{\xi}_i + 
\frac{\partial^2 {\bf r}} {\partial t^2} ,
\end{equation}
where the dot denotes a total derivative with respect to time. The
coronal plasma surrounding the blob is assumed to move with the
expanding flux rope ($\xi_{\rm cor}$ = constant), so the coronal
plasma velocity is
\begin{equation}
{\bf v}_{\rm cor} ({\bf r}_i) = \frac{\partial {\bf r}} {\partial t} .
\end{equation}
Therefore, the blob moves relative to its local surroundings and may
experience a ``frictional'' force due to its motion along the field
line (as described by $\dot{\xi}_i$). Then the equation of motion of a
blob along its field line is given by
\begin{equation}
\ddot{\bf r}_i \cdot \hat{\bf s}_i = {\bf g} ({\bf r}_i) \cdot
\hat{\bf s}_i - \beta \left[ {\bf v}_i - {\bf v}_{\rm cor} ({\bf r}_i)
\right] \cdot \hat{\bf s}_i , 
\end{equation}
where $\hat{\bf s}_i$ is the unit vector along the field line,
${\bf g} ({\bf r})$ is the acceleration of gravity, $\beta$ is a
friction coefficient, and we neglect the effects of plasma
pressure. Using $\hat{\bf s}_i = ( \partial {\bf r} / \partial \xi) /
| \partial {\bf r} / \partial \xi |$, we obtain the following
second-order differential equation for the coordinate $\xi_i (t)$:
\begin{equation}
\ddot{\xi}_i  + a_i ( \dot{\xi}_i )^2 + 2 b_i \dot{\xi}_i + c_i =
- \beta \dot{\xi}_i , \label{eq:motion}
\end{equation}
where
\begin{eqnarray}
a_i & \equiv & \left( \frac{\partial {\bf r}} {\partial \xi} \cdot
\frac{\partial^2 {\bf r}} {\partial \xi^2} \right) /
\left| \frac{\partial {\bf r}} {\partial \xi} \right|^2 , \\
b_i & \equiv & \left( \frac{\partial {\bf r}} {\partial \xi} \cdot
\frac{\partial^2 {\bf r}} {\partial \xi \partial t} \right) /
\left| \frac{\partial {\bf r}} {\partial \xi} \right|^2, \\
c_i & \equiv & \left[ \frac{\partial {\bf r}} {\partial \xi} \cdot
\left( \frac{\partial^2 {\bf r}} {\partial t^2} - {\bf g} \right) 
\right] / \left| \frac{\partial {\bf r}} {\partial \xi} \right|^2.
\end{eqnarray}
We assume that the blobs are initially at rest in the dips of the
helical field lines, $\dot{\xi}_i (0) = 0$ and ${\bf g} ({\bf r}_i)
\perp \hat{\bf s}_i$. The later condition determines the values of $p_i$ and $q_i$
describing the field lines on which the blobs are located. Using these
values, equation (\ref{eq:motion}) is numerically integrated to obtain
$\xi_i(t)$ for each blob. The positions ${\bf r}_i (t)$ and velocities
${\bf v}_i (t)$ can then be computed.

One of the free parameters of the flux rope is the number of helical windings. In our modeling
for Section 3.3, we find that the ability of the flux rope to carry matter outward depends strongly on the 
degree of twist of the flux rope. The number of turns is chosen such that at least
half of the blobs are ejected. This required about 3 turns. Unfortunately, the observations do not put
strong constraints on this number. For the model of Section 3.2 we use fewer turns because
(1) only shorter section of the flux rope is modeled, and (2) the starting time is earlier, so fewer turns
have built up.

We find that in order to slow down the speed of falling blobs we have to introduce a friction between
the blobs and the surrounding corona. Also, in the model of the rolling motion (Section 3.2) we need friction
to prevent the blobs from sliding along the field lines. The nature of this friction is unclear. One possibility
is that the pressure of the coronal plasma on either side of a blob provides a net force on the blob as 
it moves along its field line \citep{2012ApJ...745..152A}. Another possibility is that the motion of the blob causes the emission of
MHD waves that propagate out into the surrounding medium. These waves carry energy away from the blob,
so the kinetic of the blob must be reduced.

\subsection{Modeling of the Rolling Motion of Q1 and Q2 Observed by STEREO$\_$B}
\label{roll}

In the early phase of the event, the prominence Q1 and Q2
rises slowly and displays a clockwise rotating motion around its own axis 
as discussed in Section 2.2 and shown in Figure 6. Here we present a numerical
model for the observed motions. 

We assume that twist is continuously injected into the magnetic flux rope as a result of 
asymmetric reconnection occurring within the active region (see Section 4). 
Theoretical modeling of magnetic reconnection in non-symmetric configurations has shown that
asymmetries can lead to rotating motions of the ejected plasma
\citep{2012ApJ...751...56M}. The twist is assumed to propagate along the flux rope
in the eastern direction, and gives rise to the observed rolling motion. 
The present model is an empirical description of the observed
motions, and does not depend on the details of the process by which
the twist is generated.

Model 1 describes the dynamics of the filament flux rope during the
period from 14:00 UT to 16:00 UT. The parameters of this model are
listed in Table 2, and a detailed description of the meaning of these
parameters is given in the Appendix. In the initial state, a
right-helical flux rope lies horizontally above the solar surface.
Its total length corresponds to a heliocentric angle of about
$44^\circ$. The axis of the flux rope lies at a height $c_0 = 0.05$
$\rm R_\odot$, equal to the radius of cross-section ($R_0 = 0.05$
$\rm R_\odot$), hence the lower edge of the flux rope touches the
photosphere. The flux rope is centered at latitude $-35^\circ$ and
longitude $-40^\circ$ as seen from Earth. The axis of the flux rope is
tilted with respect to a line of constant latitude by an angle of
$-35^\circ$. Initially, the helical field lines make one full turn
around the axis.

Figure \ref{fig9}(a) shows the initial state of the flux rope and
filament projected onto the plane of the sky as seen from STEREO$\_$B
at 14:00 UT. The circular arc indicates the south-west quadrant of the
solar limb, and the black curves show three field lines at the outer
edge of the flux rope. The colored line segments simulate prominence
threads located in the lower half of the flux rope. The threads are
radially oriented, and each thread consist of 80 blobs that are too
close to be distinguished in this figure. Initially, each blob is
located at a dip in its field line.

Figure \ref{fig9}(b) shows the magnetic configuration at 16:00 UT
after additional twist has been injected into the flux rope from the
western side (right-hand side of the figure). The number of helical
windings has increased from 1.0 to 1.8. The field lines are held
fixed on the left-hand side ($\xi_0 = -1$), therefore, as more twist
is injected from the right the field-line dips move to the left in the
figure. We simulated the dynamics of the prominence blobs for
different values of the parameter $\beta$ describing the frictional
coupling between the blobs and their surroundings. We found that in
the case without frictional coupling ($\beta = 0$) the blobs have a
strong tendency to slide down along the magnetic field lines and
remain close to the field-line dips. However, when the coupling
constant $\beta$ is large, the motions of the blobs along the field
lines are suppressed, and the blobs tend to follow the rotating
motion of the flux rope. The latter leads to better agreement with the
STEREO$\_$B observations.  Figure \ref{fig9}(b) shows the results for
$\beta = 0.01$ $\rm s^{-1}$, which corresponds to a frictional time
scale of only 100 s, much shorter that the time scale for blobs to
slide down along the field lines. Note that the threads on the western
end of the filament have rotated in the clockwise direction,
consistent with the STEREO$\_$B observations. This rotation can also be
seen in a video (video4) of the simulation results available in the on-line
version of the paper. To better visualize the rotation, the blobs are
colored according to their line-of-sight (LOS) velocities: blue and
red for motions toward and away from the observer, and magenta for
zero velocity. We conclude that a relatively strong coupling between
the prominence blobs and their surroundings is needed to reproduce the
observations.

\subsection{Main Eruption}

\label{main}

In this section we present a numerical model of the main eruption of filaments AR1, AR2, Q1, Q2, and Q3 on December 6.
The STEREO observations indicate that the part of the filament overlying the quiet Sun begins to move radially outward at about 16:00
UT, while the part overlying the active region remains anchored until
about 18:00 UT. Model 2 describes the main eruption  in a simplified 
way, and we take the starting time to be 16:00 UT.
The parameters of the model are listed in Table 2 (for details, see
the Appendix). In the initial state of the model, a right-helical
flux rope lies curved initially above the solar surface, and its
total length corresponds to a heliocentric angle of about $63^\circ$.
The axis of the flux rope is tilted by $-10^\circ$ relative to a line
of constant latitude. The helical field lines initially make three
full turns around the axis. These parameters are different from those
of Model 1 because we now intend to describe the large-scale structure
of the flux rope.

Figures \ref{fig10}a--\ref{fig10}c show the initial state of the flux rope as seen
from SDO,  STEREO$\_$B and STEREO$\_$A at 16:00 UT. 
The circular arc indicates the south-east quadrant of the solar limb. 
The magenta segments simulate prominence
threads located in the lower half of the helical flux rope. The
threads are nearly vertically oriented, but with a small tilt towards
the south, consistent with the AIA observations of this prominence
(see Paper~I). Each thread consist of 80 individual blobs that are
initially too close together to be distinguished. The blobs are
located at dips in the field lines.

During the eruption the center of the flux rope moves radially outward
in the corona, and the legs become more vertical. Figures \ref{fig10}d--\ref{fig10}f
shows the magnetic configuration as seen
from SDO,  STEREO$\_$B and STEREO$\_$A at 17:36 UT. 
Three corresponding videos (videos 5--7) showing the evolution from 16:00 UT to 19:00 UT are available in the on-line
version of the paper. The model shows that near the top of the
flux rope the blobs are carried radially outward by the dips of the
field lines. However, in the legs the
dips soon disappear and the blobs start falling back down to the
photosphere. The speed at which the blobs fall down depends strongly
on the value of the friction parameter $\beta$. This parameter was
chosen by comparing the modeling results with the observed downflow
velocities as seen with AIA. The observed speeds typically reach about
100 $\rm km ~ s^{-1}$, which can be reproduced with $\beta \sim 0.001$
$\rm s^{-1}$. We adopt this value for Model 2. However, it should be
noted that the model provides only a very crude fit to the
observations, so the value of $\beta$ is highly uncertain.

Figure \ref{fig10} (and the associated videos) shows that for a given
thread in the leg of the flux rope the blobs at the top start to fall
first, followed later by the blobs at the bottom of that thread. This
is due to the fact that the dips are removed from the top down.
Therefore, as they fall the simulated threads are turned upside down.
This rotating motion of the threads (left leg in Figure \ref{fig10}d) is consistent with the
observations (black arrows in Figure \ref{fig7}c) discussed in Section 2.3. We conclude that the observed
motions of the threads are consistent with the existence of dips in
the field lines in the pre-eruption state, and the subsequent removal of those
dips from the top down during the eruption.

A comparison of Figures \ref{fig10}d--\ref{fig10}f and
Figures \ref{fig2}c, \ref{fig3}c, and \ref{fig4}c suggests that the modeled overall shape of the erupting
prominence match the observations. However, there are some differences in the evolution of the right leg
between observations and the model. In observations, the right prominence leg erupts much later than the 
other part of the prominence, which is possibly due to the strong overlying magnetic field in the active region.
In the model the entire flux rope erupts radially outward at the same time.

\section{Origin of the Rolling Motion of Prominence Q1 and Q2}

In Section 3.2, we presented a rolling motion of Prominence Q1 and Q2 that occurs at the early phase of the eruption on December 6 from 14:21 UT to 17:00 UT.  STEREO$\_$B 195~\AA~observations show that the filament sheet is rotating around its own axis in the clockwise direction as viewed from the east to west. The motion starts from the active region, then propagates towards the east. This motion is similar to the roll effect identified by \citet{2008ASPC..383..243P} in the sense that it is a rotation around the axis of the filament. In all cases of the roll effect recognized to date, there has been a one-to-one relationship between the chirality of the filament and the direction of the roll with dextral filaments always rolling toward the positive photospheric magnetic field side of the prominence and sinistral filaments rolling in the opposite direction \citep{2003AdSpR..32.1883M, 2008ASPC..383..243P, 2011JASTP..73.1129P}. However, the filament studied in the current paper is sinistral (also see Paper~I) and displays rolling motion towards the positive photospheric magnetic field side of the prominence.

There are several candidate mechanisms for the observed rolling motions: (1) untwisting of the magnetic field in the rising flux rope during expansion and relaxation; (2) local magnetic force imbalance caused by the presence of a coronal hole near the filament channel \citep{2011JASTP..73.1129P}; (3) increase of twist due to reconnection below a flux rope or sheared arcade; (4) asymmetric reconnection due to an offset between the CME current sheet and the rising flux rope \citep{2012ApJ...751...56M}. In the present case, the erupting filament has a sinistral orientation of its axial field, and has right-helical (positive) twist. The observed clockwise rolling motion indicates an increase of twist rather than untwisting of the magnetic field, which rules out the first mechanism. Moreover, our simulation suggests that a continuous increase of twist in the flux rope is consistent with the observed rolling motion. The second mechanism cannot be ruled out because there is indeed a coronal hole to the north of the active region. However, it seems unlikely that the weak fields of the coronal hole can affect the flux rope in the active region at the early slow rising phase of the eruption.

The third mechanism can lead to twist increase in the existing flux rope or sheared arcade. There are two alternatives. First, reconnection (beneath the flux rope) of weakly sheared flux that is rooted immediately outside the filament channel would yield a highly twisted layer surrounding the existing, weakly twisted flux rope. Second, reconnection beneath a sheared arcade will convert shear into twist, yielding a flux rope in which the number of twist is dictated by the shear length and the filament channel length. However, it is difficult to imagine how this twist increase will affect the preexisting filament material located near the center of the flux rope or sheared arcade. 
In particular, it is unclear whether such reconnection can cause rolling motion of the existing filament. In the following, we focus on the fourth mechanism, i.e., we suggest that
the observed rolling motion may be due to asymmetric reconnection. 

Figure \ref{fig11} shows cross sections of the distribution of field-aligned electric currents along different parts of the flux rope at the active region from one non-linear force-free field model (NLFFF, Model 1) presented in Paper~I. The currents are characterized by the $\alpha$ parameter of the NLFFF
($\nabla\times\textbf{B}=\alpha\textbf{B}$). The background images in the top row are the maps of the radial component of the photospheric magnetic field observed by SDO/HMI. The zero point of the longitude corresponds to the central meridian on 2010 December 10 at 14:00 UT. The blue curve refers to the path where the flux rope is inserted. The bottom row shows the distribution of $\alpha$ in vertical cross-sections along different parts of the flux rope (as indicated by the yellow line in the corresponding top-row image) from the NLFFF model. The zero points of the X-axis in the $\alpha$ plots are the northern ends of the yellow lines in the top images. The top row of Figure \ref{fig11} shows that the distribution of the magnetic fields on the two sides of the PIL (blue line) is asymmetric on the east and west sides of the active region as marked by the two yellow lines in Figures \ref{fig11}a and \ref{fig11}c. However, in the middle of the active region the magnetic fields are nearly symmetric on the two sides of the PIL (yellow line in Figure \ref{fig11}b). The bottom row shows that the $\alpha$ distribution is also asymmetric on two sides of the active region (Figures \ref{fig11}d and \ref{fig11}f), while it is nearly symmetric in the middle of the region (Figure \ref{fig11}e). The $\alpha$ distribution on the east side of the active region is inclined towards the south, while on the west side of the region, the $\alpha$ distribution is inclined towards the north.


A scenario for the generation of rolling motions and the formation of magnetic twist along the flux rope at and near the active region is presented in Figure \ref{fig12}. Before the eruption, the active region contains a flux rope with weakly positive twist. The axial flux in the flux rope points towards the left when viewed from the positive-polarity side, see Figure \ref{fig12}a. After the onset of reconnection in the middle part of the active region, reconnection propagates towards the east and west sides of the region. The asymmetry of the $\alpha$ distribution in the NLFFF model suggests the existence of an offset between the vertical current sheet and the slowly rising flux rope as shown in Figure \ref{fig12}b. The asymmetric reconnection induced by this offset may result in the rolling motions as shown by the dashed arrows in Figures \ref{fig12}a and \ref{fig12}b. 
Note that the directions of the rolling motions are predicted to be opposite in the eastern and western sides of the active region. The filament in the active region should not roll since the underlying current sheet is not offset. Here we assume that the flux rope is held fixed at the two ends. However, the different rolling motions will lead to the twist increase on the two sides of the active region, while reduce the twist or even reverse the sign of the twist in the active region. One possible result is the formation of magnetic twist that varies along the flux rope, as shown in Figure \ref{fig12}c. This scenario provides a natural explanation on the observed rolling motion on the eastern side of the active region. This alternate twist along the flux rope exists as long as the ``driving force" caused by asymmetric reconnection exists. Once the ``driving force" stops, the twist will propagate until it is uniform throughout the flux rope. 

\citet{Thompson2012} determines magnetic twists of the filament during the eruption on December 6 using triangulations of prominence threads observed by STEREO. \citet{Thompson2012} finds that the prominence threads display negative twist in the active region (his Figure 11), while positive twist is identified for the eastern and western sides of the filament (his Figures 9--10). This result is consistent with our observed rolling motion and interpretation as shown in Figure 12 in the present paper.

\section{Conclusions}

We present observations and modeling of the eruption of a large polar crown prominence on 2010 December 6. This complex prominence is composed of different segments. The middle part of the prominence is located in the active region remnant. Observations suggest that this middle part contains two layers, one layer is elevated 
above the active region, the other layer is located low at the core of the region. The majority of the prominence is located on the eastern (left) side of the active region, while a small part of the filament is located on the western (right) side of the region. The primary focus of the current study is the eastern side of the quiescent filament,
which is divided into four parts according to the different behavior during the eruption. Q1 filament is located right next to the active region, while Q2, Q3, and E4 filament segments are located sequentially further away from the active region.

The eruption begins with appearance of brightenings immediately surrounding the active region filament at 14:16 UT on December 6 as observed by AIA. STEREO$\_$B observes that the filament in the core of the active region breaks apart around 14:21 UT. Part of the filament ejects towards the east which leads to the explosive eruption of the eastern (left) part of the quiescent prominence, which is the focus of the present paper. The other part of the filament material ejects to the west which leads to a clear rise of the western (right) end of the filament followed by an eruption at 2:06 UT on December 7. Though the instability appears to begin in the core of the active region, the elevated filament above the active region remains stable until the eastern quiescent filament rises to a much larger height. The prominence eruption on December 6 appears to be a sideways/non-radial eruption in which the filament material escapes from the weaker field region firstly. 

Thanks to the three views of the Sun by STEREO and SDO, we have identified two types of rotating motions during  the prominence eruption on 2010~ December 6. To understand the observed rotating motions from different points of view, we develop a simplified three-dimensional model. The prominence is assumed to consist of a collection of discrete blobs that are tied to particular field lines. The time-varying magnetic field is described by a simple analytical model of a helical flux rope, and the motion of the blobs is determined by solving the equations of motion along the field lines, taking into account the effects of gravity and friction with the surrounding corona.

The first rotating motion occurs at the early phase of the (from 14:21 UT to 17:00 UT) eruption of the eastern side quiescent filament (Q1 and Q2). STEREO$\_$B 195~\AA~observations show that the filament sheet is rotating around its own axis in the clockwise direction as viewed from the east to west. This rolling motion is not clearly visible in SDO and STEREO$\_$A observations. The motion starts from the active region, then propagates towards the east. The observed rolling motion is reproduced by continuous twist injection into the flux rope in Model 1 from the active region side. We suggest that the observed rolling motion is most likely caused by asymmetric reconnection induced by the asymmetric distribution of the magnetic fields on the two sides of the filament. The second type of rotating motion is observed by SDO, but is not visible in STEREO observations. AIA 304~\AA~observations suggest that Q3 filament makes a half turn during its process of rising and draining back towards the Sun. The northern leg of the Q2 filament also displays a similar rotating motion, though not as remarkable as Q3 filament. This rotating motion is very well reproduced by our Model 2. 
Our model suggests that this rotating motion of prominence threads is consistent with the existence of dips in the field lines in the pre-eruption state,
and the removal of those dips from the top down during the eruption.

Our results suggest that the following scenario for the event: (1) Reconnection in the active region  leads to twist increase in the existing flux rope (or enabled the formation of a flux rope). (2) The twist on the quiet Sun reached a critical threshold for kink instability, leading to an eruption of the eastern quiescent filament. (3) Later on the elevated layer of the active region filament is pulled out as well. Both our model result and the observational result by \citet{Thompson2012} indicate that there is no strong writhing motion during the eruption.

Acknowledgments: We acknowledge the anonymous referee for valuable comments to improve the paper. Hinode is a Japanese mission developed and launched by ISAS/JAXA, with NAOJ as domestic partner and NASA and STFC (UK) as international partners. It is operated by these agencies in co-operation with ESA and the NSC (Norway). We thank the team of SDO/AIA, SDO/HMI, STEREO/EUVI, Hinode/XRT for providing the valuable data. The STEREO and HMI data are downloaded via the Virtual Solar Observatory and the Joint Science Operations Center. This project is partially supported under contract NNM07AB07C from NASA to the Smithsonian Astrophysical Observatory (SAO) and SP02H1701R from LMSAL to SAO as well as NASA grant NNX12AI30G.

\appendix

\begin{center} {\bf APPENDIX} \end{center}

\section{Analytic Model for an Erupting Flux Rope}

The prominence plasma is assumed to be located in a curved, helical
flux rope that expands with time. Here we describe the time-dependent
shapes of the field lines inside the flux rope. Each field line
follows a curved path ${\bf r} (\xi,t)$, where $\xi$ measures position
along the field line, and $t$ is the time as measured from the start
of the eruption. The parameter $\xi$ varies from $-1$ at one endpoint
of the flux rope to $+1$ at the other end. Positions are described in
terms of a Cartesian coordinate system $(x,y,z)$ with the origin
located at Sun center and the $+z$ axis along the direction of
flux-rope expansion. All lengths are given in units of the solar
radius.

We first consider the shape of the flux-rope axis, which is a special
field line at the center of the flux rope. Its non-circular shape
${\bf r}_{\rm axis} (\xi,t)$ is given by the following expressions:
\begin{eqnarray}
x_{\rm axis} (\xi ,t) & = & r_0 \sin ( \theta \xi) /
\cos^n (a_0 \theta \xi) , \label{eq:xaxis} \\
y_{\rm axis} (\xi, t) & = & w_0 \sin ( \pi \xi ) ,
\label{eq:yaxis} \\
z_{\rm axis} (\xi ,t) & = & r_0 \cos ( \theta \xi) /
\cos^n (a_0 \theta \xi) + z_0 , \label{eq:zaxis} 
\end{eqnarray}
where $\theta(t)$ is the ``angular width'' of the curved path,
$r_0(t)$ is the radius of curvature at the top of the path,
$z_0(t)$ is the position of the curvature center, and $w_0(t)$
describes the writhe of flux-rope axis. 
The parameter $a_0$ is a constant that determines the width-to-height
ratio of the expanding flux rope ($0 < a_0 < \onehalf$), and the
exponent $n \equiv 1/a_0$. During the eruption, the middle part of the
flux rope rises, but the endpoints are assumed to remain at fixed
positions in the low corona: $x_{\rm axis} (\pm 1,t) = \pm d_0$ and
$z_{\rm axis} (\pm 1,t) = h_0$, where $d_0$ and $h_0$ are constants.
It follows that the functions $r_0(t)$ and $z_0(t)$ are given by
\begin{eqnarray}
r_0(t) & = & d_0 \cos^n (a_0 \theta) / \sin \theta , \label{eq:r0} \\
z_0(t) & = & h_0 - d_0 / \tan \theta . \label{eq:z0} 
\end{eqnarray}
The constant $h_0$ is chosen such that the endpoints lie at a fixed
height $R_0$ above the solar surface:
\begin{equation}
h_0 = \sqrt{(1+R_0)^2 - d_0^2} , \label{eq:h0}
\end{equation}
where $R_0$ is the radius of the cross-section of the flux rope at the
two endpoints. The unit vector $\hat{\bf s}_0 (\xi,t)$ along the flux
rope axis can be found by differentiating expressions
(\ref{eq:xaxis}), (\ref{eq:yaxis}) and (\ref{eq:zaxis}):
\begin{equation}
\hat{\bf s}_0 \equiv \frac{\partial {\bf r}_{\rm axis} / \partial \xi}
{| \partial {\bf r}_{\rm axis} / \partial \xi |} =
\cos (a_1 \theta \xi) \cos \gamma ~ \hat{\bf x} 
+ \sin \gamma ~ \hat{\bf y}
- \sin (a_1 \theta \xi) \cos \gamma ~ \hat{\bf z},
\label{eq:s0}
\end{equation}
where $a_1 \equiv 1 - a_0$, and the angle $\gamma (\xi,t)$ is defined
by
\begin{equation}
\tan \gamma= \frac{\pi w_0}{\theta r_0} \cos^n (a_0 \theta \xi)
\cos ( \pi \xi ) . \label{eq:gamma}
\end{equation}
The outward motion of the flux rope axis is given by
\begin{equation}
z_{\rm top} (t) \equiv z_{\rm axis} (0,t) = 1 + c_0 +
v_{\rm max} [ t - \tau_{\rm acc} ( 1 - e^{-t/\tau_{\rm acc}} ) ] ,
\label{eq:ztop}  
\end{equation}
where $\tau_{\rm acc}$ is the time scale of the initial outward
acceleration of the flux rope, $v_{\rm max}$ is its terminal
velocity, and $c_0$ is the initial height of the flux rope axis
above the solar surface ($z_{\rm top}(0) = 1 + c_0$). The function
$\theta(t)$ is computed by inverting the equation $z_{\rm top} (t) =
d_0 f[\theta(t)] + h_0$, where $f(\theta) \equiv [ \cos^n (a_0 \theta)
- \cos \theta ] / \sin \theta$, which follows from equation
(\ref{eq:zaxis}). The out-of-plane motion of the axis is described by
\begin{equation}
w_0 (t) = v_{\rm writhe} [ t - \tau_{\rm writhe}
( 1 - e^{-t/\tau_{\rm writhe}} ) ] , \label{eq:w0}  
\end{equation}
where $\tau_{\rm writhe}$ is the time scale for the onset of the
writhing motion, and $v_{\rm writhe}$ is its terminal velocity.

The flux rope is assumed to have a circular cross-section. The radius
$R(\xi,t)$ of the cross section varies with time $t$ and with position
$\xi$ along the flux rope:
\begin{equation}
R(\xi,t) = R_0 \left\{ \frac{1 - \xi^2}{\cos [\theta(t)/2]}
+ \xi^2 \right\} .
\end{equation}
For the model describing the main eruption (section XXX) we use
$a_0 = 0.28$, $d_0 = 0.55$ $\rm R_\odot$ and $c_0 = 0.16$
$\rm R_\odot$, which yields $\theta (0) = 1.153$ rad. Therefore,
in the initial state $\cos [\theta(0)/2] = 0.838$, and the radius
of the flux rope is approximately constant along the flux rope,
$R(\xi,0) \approx R_0$. However, as time progresses the radius
$R(0,t)$ at the top of the flux rope increases significantly.

The shape of the helical field lines can now be approximated as
follows: 
\begin{eqnarray}
{\bf r} (p,q,\xi,t) & = & {\bf r}_{\rm axis} (\xi,t) + R(\xi,t)
[ p \cos \phi(\xi,t) - q \sin \phi(\xi,t) ] ~ \hat{\bf s}_1 (\xi,t)
\nonumber \\
 & & ~~~~~~~~ \cdots + R(\xi,t) 
[ p \sin \phi(\xi,t) + q \cos \phi(\xi,t) ] ~ \hat{\bf s}_2 (\xi,t)
\label{eq:rpos}
\end{eqnarray}
where $p$ and $q$ are constants along a given field line, and are also
constants of motion; they can be used as labels of the field lines.
The interior region of the flux rope is given by $p^2 + q^2 \leq
1$. The vectors $\hat{\bf s}_1 (\xi,t)$ and $\hat{\bf s}_2 (\xi,t)$
are defined by
\begin{eqnarray}
\hat{\bf s}_1 (\xi,t) & = & \sin (a_1 \theta \xi) ~ \hat{\bf x}
+ \cos (a_1 \theta \xi) ~ \hat{\bf z}, \label{eq:s1} \\
\hat{\bf s}_2 (\xi,t) & = & \cos (a_1 \theta \xi) \sin \gamma ~
\hat{\bf x} - \cos \gamma (\xi,t) ~ \hat{\bf y} 
- \sin (a_1 \theta \xi) \sin \gamma ~ \hat{\bf z} ,
\label{eq:s2}
\end{eqnarray}
so that the vectors $\hat{\bf s}_0$, $\hat{\bf s}_1$ and
$\hat{\bf s}_2$ are mutually orthogonal. Hence, the helical
displacement of the field lines is perpendicular to the flux rope
axis. The phase angle $\phi(\xi,t)$ is given by
\begin{equation}
\phi(\xi,t) \equiv \pi N(t) ( \xi - \xi_0) , \label{eq:phi}
\end{equation}
where $N(t)$ is the number of helical windings along the length of the
flux rope (i.e., the number of full turns as we follow a field line
from $\xi = -1$ to $\xi = +1$). The number of windings $N(t)$ is
the same for all field lines within the flux rope, and does not vary
with radial distance from the flux rope axis. $N(t)$ is positive
(negative) for a right-handed (left-handed) helix, and is assumed to
be a linear function of time:
\begin{equation}
N(t) = N_0 + N_1 t , \label{eq:Nt}
\end{equation}
where $N_1$ is the rate of change of the number of helical windings.
The effect of $N_1$ is to produce rotating motions inside the flux
rope. The parameter $\xi_0$ in equation (\ref{eq:phi}) determines the
position along the flux rope where there is no rotating motion.
In this paper we use $\xi_0 = -1$ to keep the field lines fixed at one
of the endpoints of the flux rope.

\newpage

\begin{figure} 
\begin{center}
\epsscale{1.0} \plotone{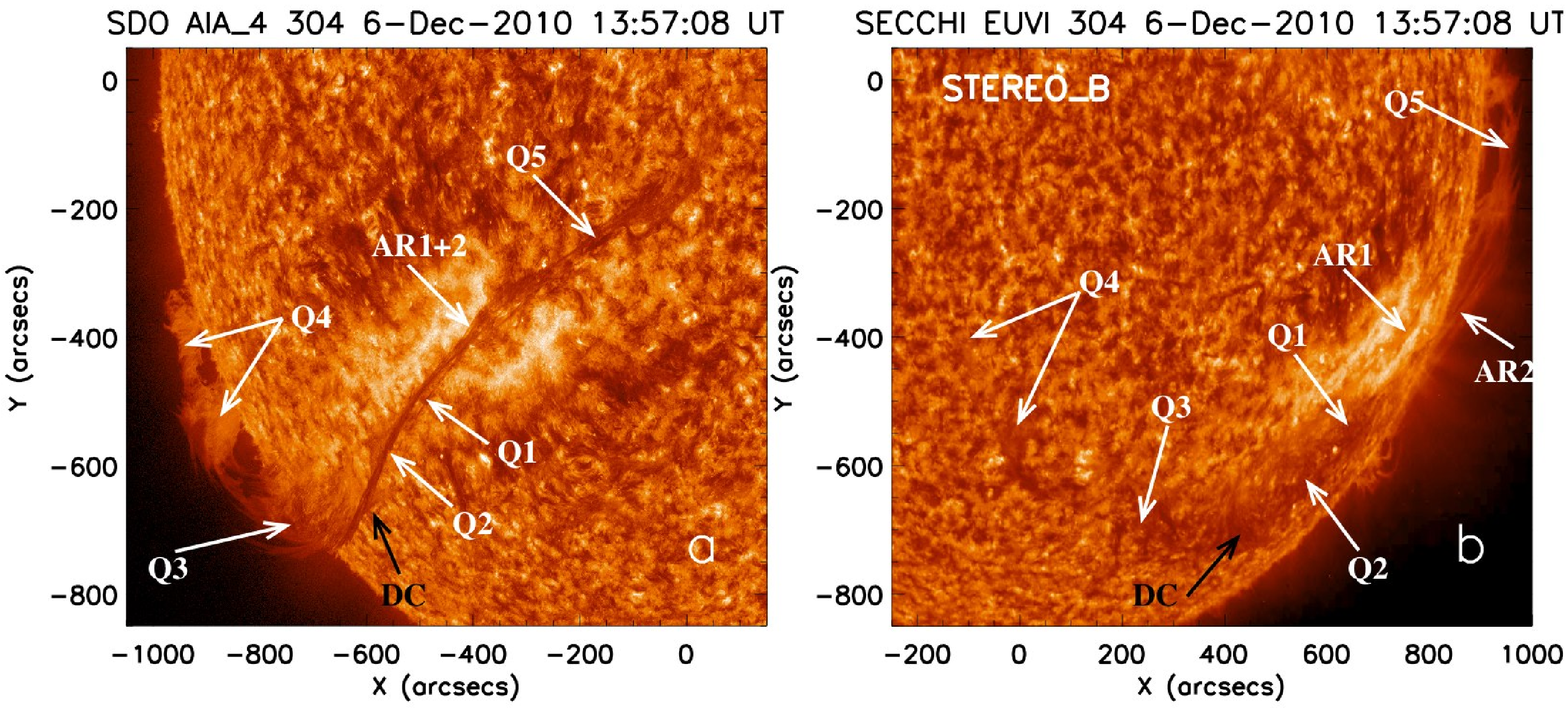}     
\end{center}
\caption {SDO/AIA (left) and STEREO$\_$B/EUVI (right) observations of the polar crown prominence at 304~\AA~before its eruption on 2010 December 6. 
This prominence is divided into 7 different segments which are marked with white arrows. The black arrows refer to the dense column (DC) structure. 
A color version of the figure is also available in the electronic edition of the \emph{Astrophysical Journal}. }
\label{fig1}
\end{figure}

\begin{figure} 
\begin{center}
\epsscale{0.7} \plotone{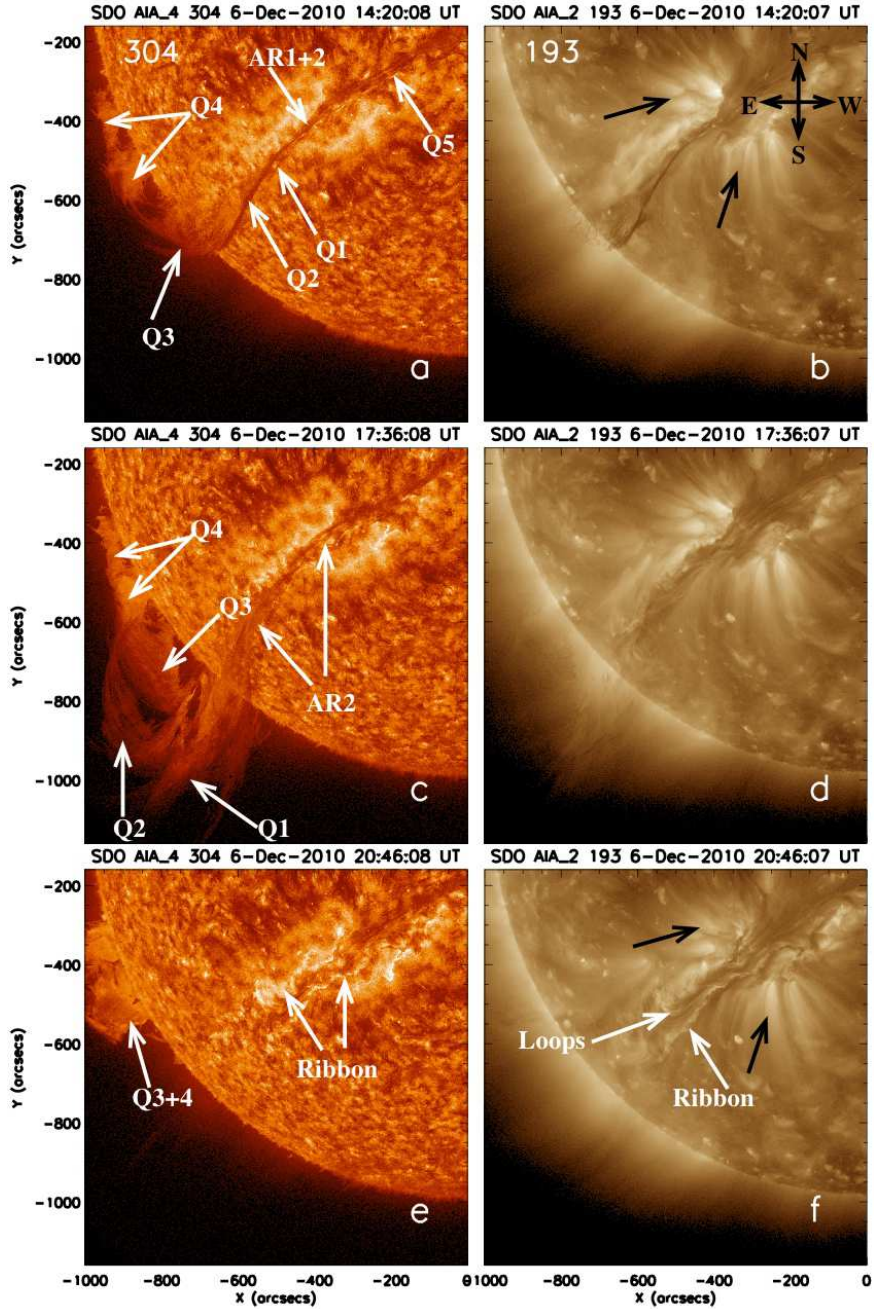}     
\end{center}
\caption {SDO/AIA observations of the prominence eruption on 2010 December 6. The images in the left and right columns are taken at 304~\AA~and 193~\AA, respectively. 
Different segments of the prominence are marked with white arrows. The black arrows refer to the straight features on the two side of the active region filament channel.
A color version of the figure and a video (video 1) of this event observed by AIA are also available in the electronic edition of the \emph{Astrophysical Journal}. }
\label{fig2}
\end{figure}

\begin{figure} 
\begin{center}
\epsscale{0.7} \plotone{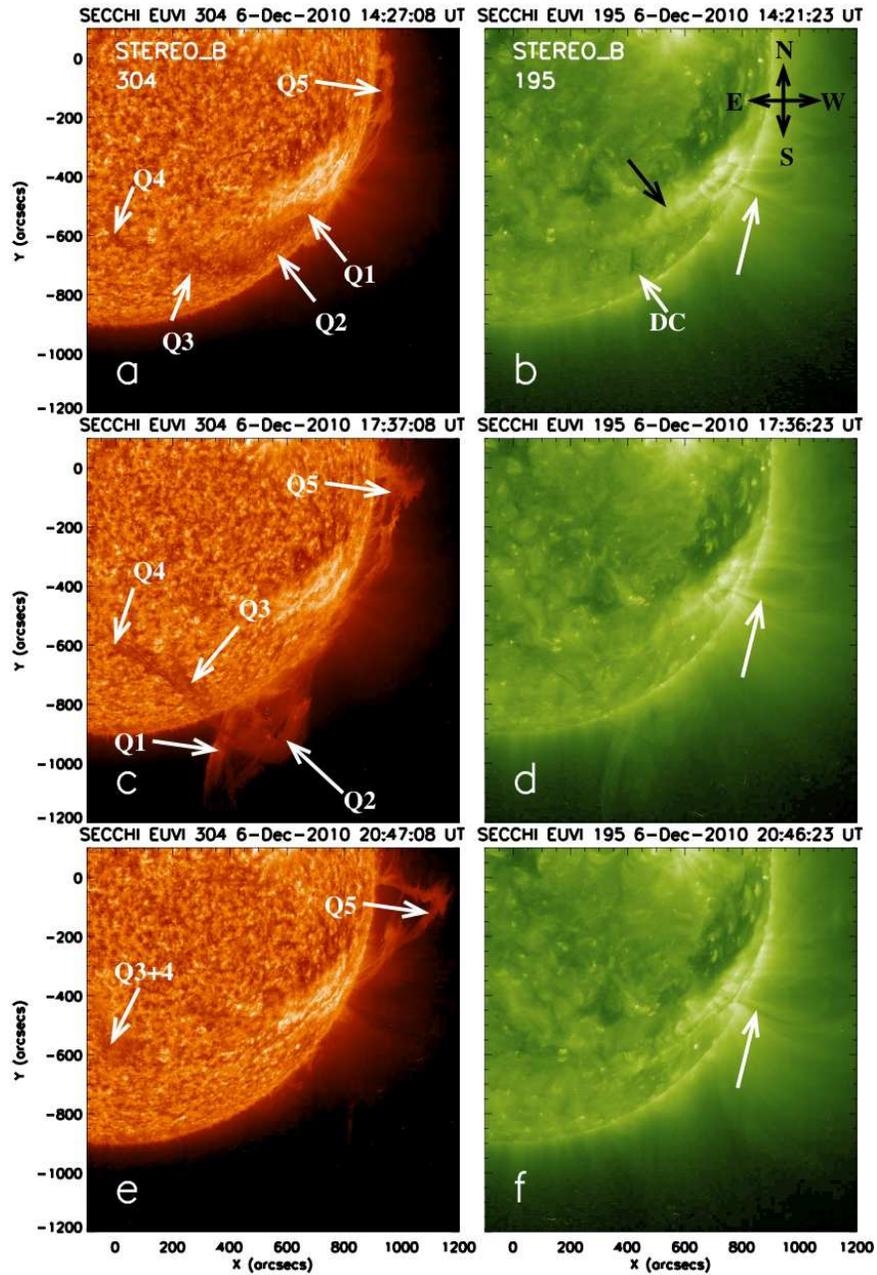}     
\end{center}
\caption{STEREO$\_$B/EUVI observations of the prominence eruption on 2010 December 6. The left and right columns show images taken at 304~\AA~and 195~\AA, respectively. Different segments of the prominence are marked with white arrows. The black arrows refer to the bright curved structure on the northern side of the quiescent filament channel. A color version of the figure and a video (video 2) of this event observed by STEREO$\_$B/EUVI are also available in the electronic edition of the \emph{Astrophysical Journal}.}
\label{fig3}
\end{figure}

\begin{figure} 
\begin{center}
\epsscale{0.5} \plotone{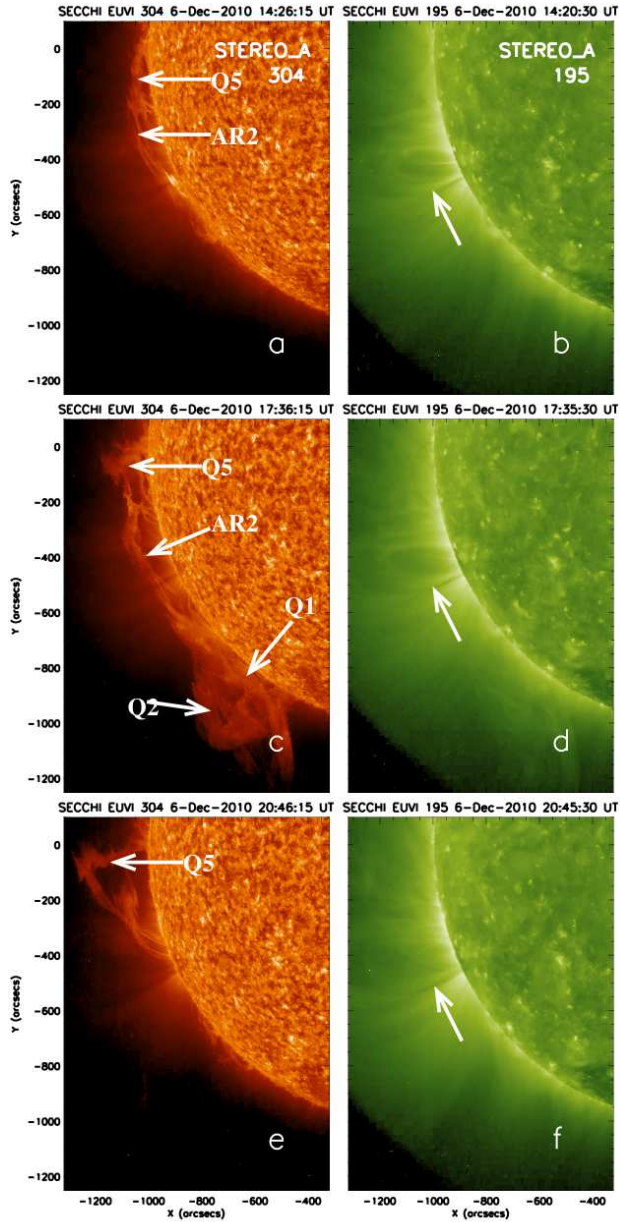}     
\end{center}
\caption{STEREO$\_$A/EUVI observations of the prominence eruption on 2010 December 6. The left and right columns show images taken at 304~\AA~and 195~\AA, respectively. The white arrows in the left and right columns refer to different segments of the prominence and overlying loops in the active region, respectively. 
A color version of the figure and a video (video 3) of this event observed by STEREO$\_$A/EUVI are also available in the electronic edition of the \emph{Astrophysical Journal}.}
\label{fig4}
\end{figure}

\begin{figure} 
\begin{center}
\epsscale{1.0} \plotone{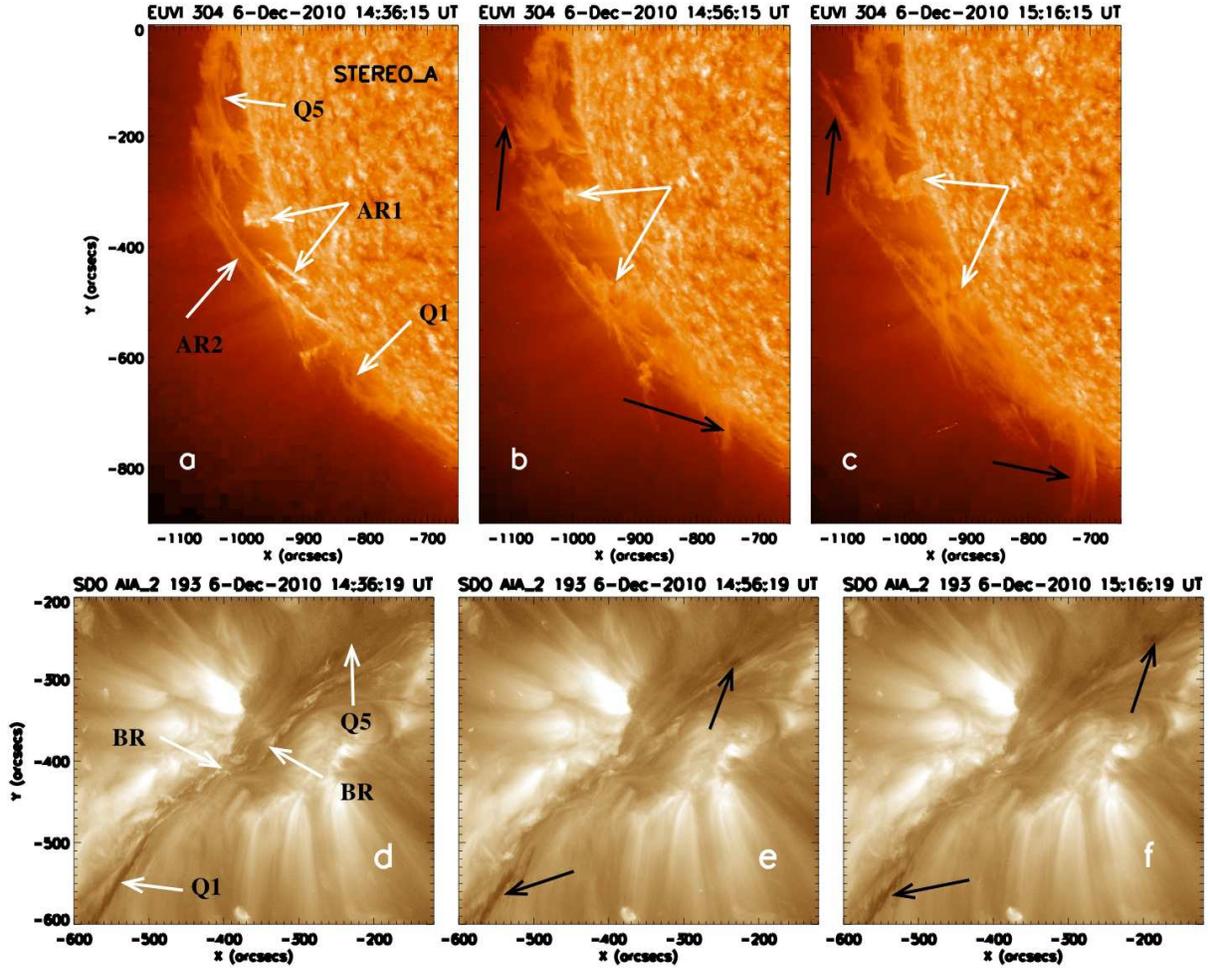}     
\end{center}
\caption{STEREO$\_$A/EUVI (304~\AA) and SDO/AIA (193~\AA) observations of the active region filament at the early stage of the eruption. Different segments of the prominence are marked with white arrows. The black arrows mark the two-direction ejection of the filament. A color version of this figure is also available in the electronic edition of the \emph{Astrophysical Journal}.}
\label{fig5}
\end{figure}

\begin{figure} 
\begin{center}
\epsscale{0.8} \plotone{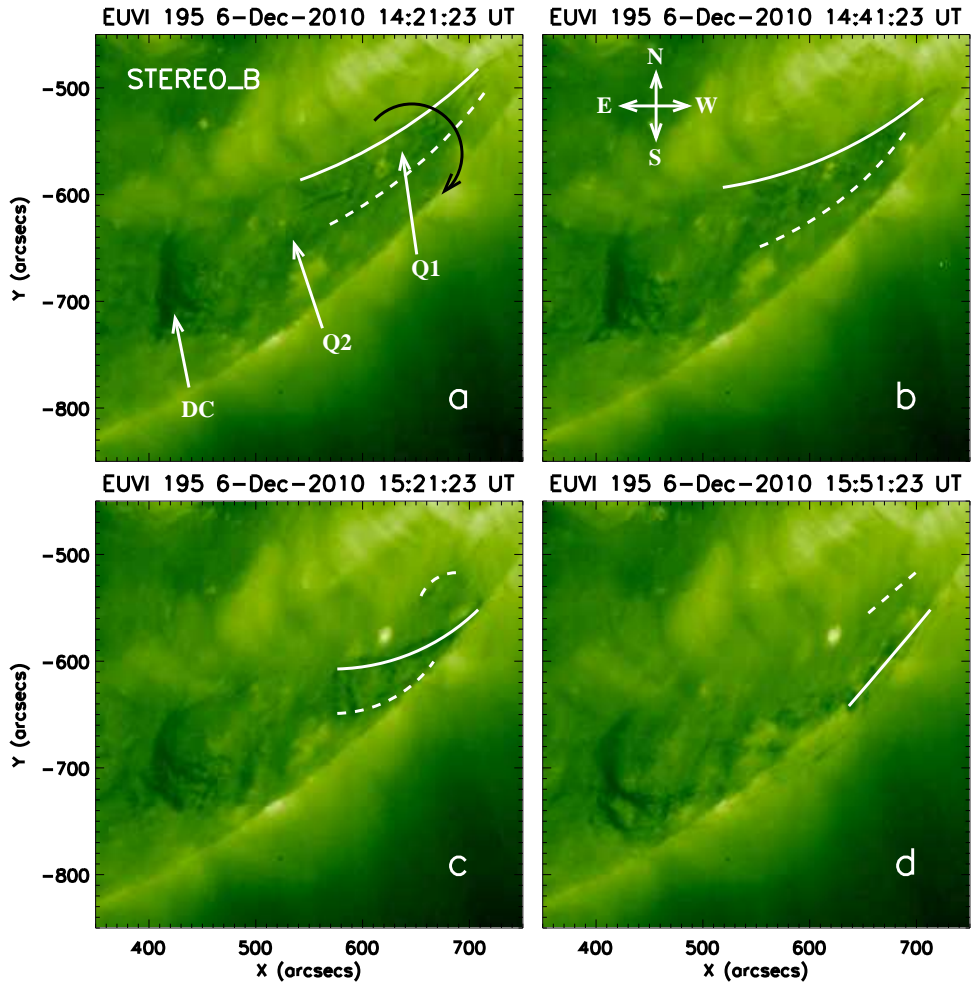}     
\end{center}
\caption {STEREO$\_$B/EUVI 195~\AA~observations of the rotating motion of the Q1 and Q2 prominences. The white solid and dashed line refer to the initial bottom and top 
border of the filament. The rotating direction of the filament is represented by the black arrow. The white arrows mark different segments of the filament as well as the dense column (DC) structure. A color version of the figure is also available in the electronic edition of the \emph{Astrophysical Journal}.}
\label{fig6}
\end{figure}

\begin{figure} 
\begin{center}
\epsscale{1} \plotone{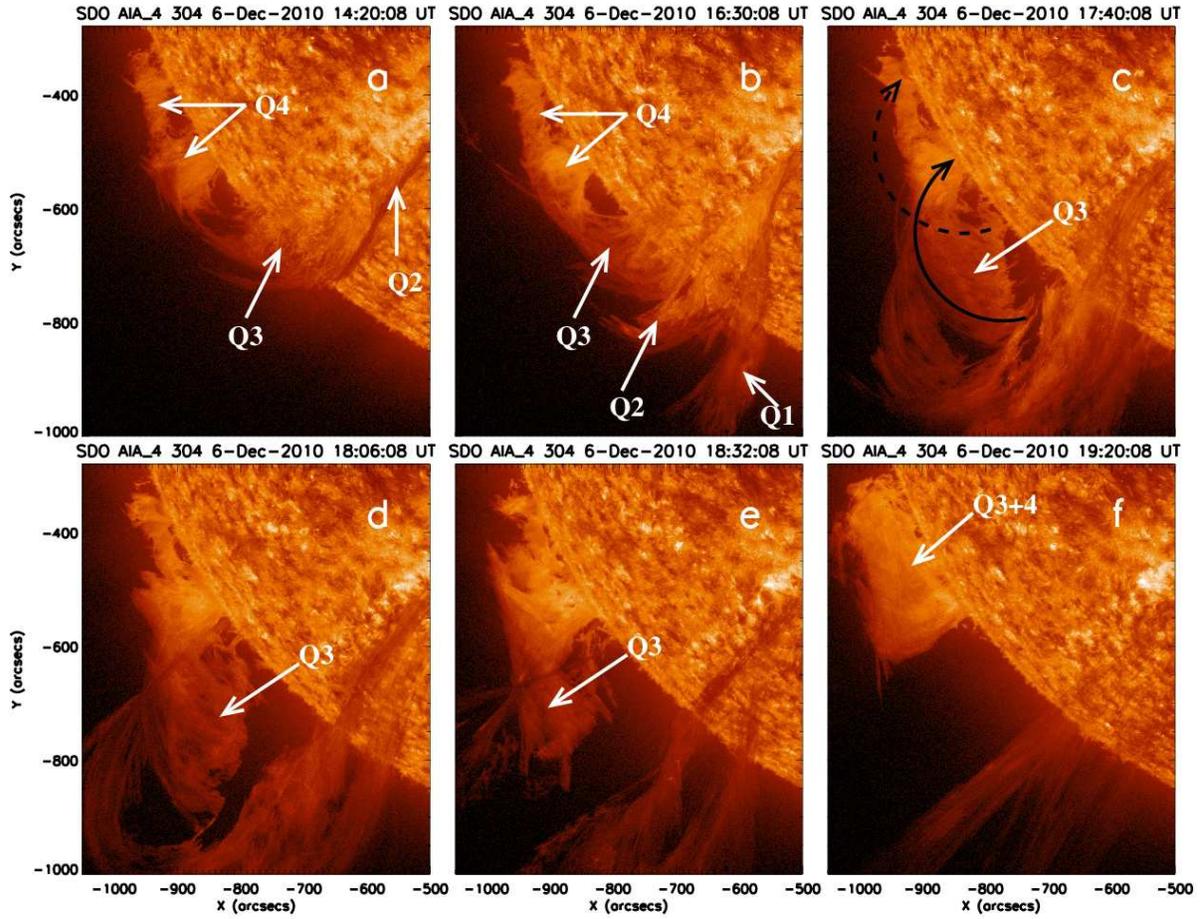}     
\end{center}
\caption {SDO/AIA 304~\AA~observations of the rotating motion of the Q3 prominence. The white arrows refer to different segments of the filament. The rotating direction of the filament is represented by the black arrow. A color version of the figure is also available in the electronic edition of the \emph{Astrophysical Journal}.}
\label{fig7}
\end{figure}

\begin{figure} 
\begin{center}
\epsscale{0.5} \plotone{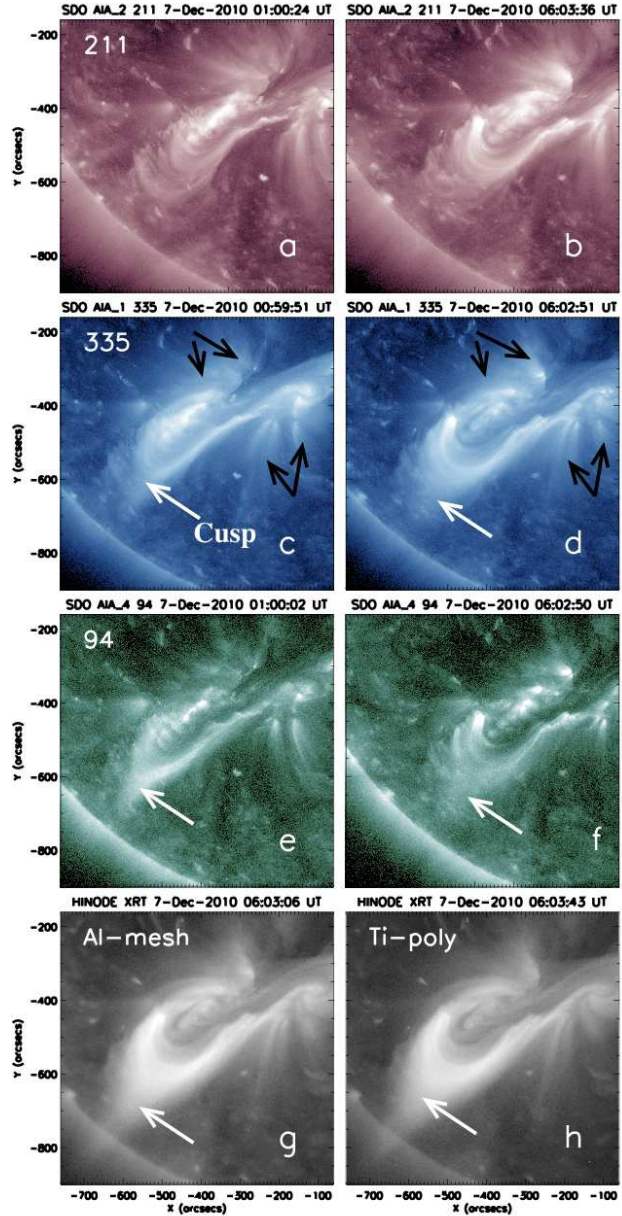}     
\end{center}
\caption{SDO/AIA and Hinode/XRT observations of the post-eruption arcades. The first (211~\AA), second (335~\AA), and third (94~\AA) rows show AIA images taken around 01:00 (left column) and 06:02 UT (right column) on 2010 December 7. The two images in the last row are taken with Al-mesh and Ti-poly filters by XRT at 06:03 UT on December 7. The AIA images at 335~\AA~and 94~\AA~are summed then averaged over 20 images. The cusp structure is marked with white arrows. The black arrows refer to the straight features on the two side of the active region filament channel. A color version of the figure is also available in the electronic edition of the \emph{Astrophysical Journal}.}
\label{fig8}
\end{figure}
 
 \begin{figure}[t]
 \begin{center}
 \epsscale{0.8}
\plotone{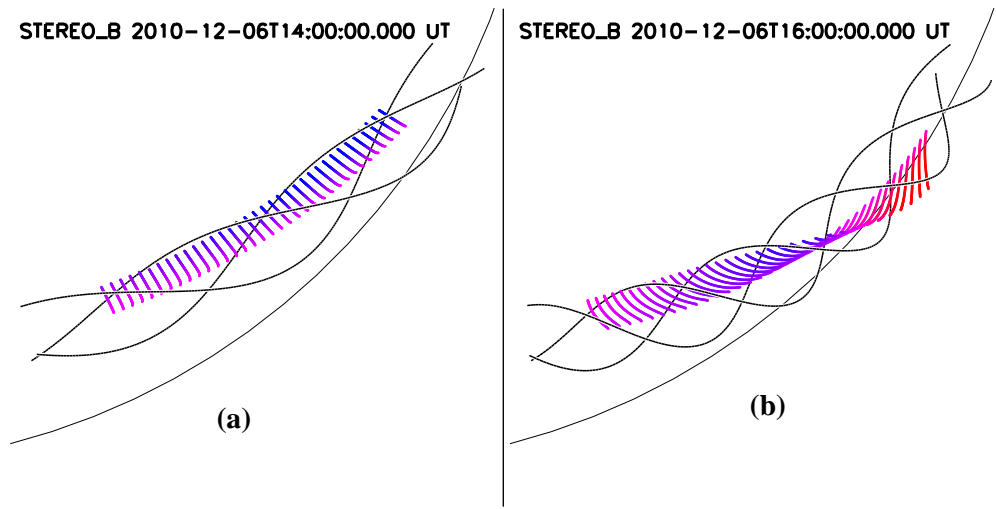}
\end{center}
\caption{Model for the rotating motion of the filament observed by
STEREO$\_$B early in 2010 December 6 eruption event. (a) Initial state at
14:00 UT. The black curves indicate helical field lines at the outer
edge of the flux rope, and the magenta line segments simulate nearly
vertical threads located in the lower half of the helical flux
rope. (b) Configuration at 16:00 UT after the prominence has been
rotated in the clockwise direction by about half a turn. The colors of
the blobs indicate the LOS velocity ($\pm 10$ $\rm km ~ s^{-1}$). An
animated version of this figure (video4) is available online.}
\label{fig9}
\end{figure}

\begin{figure}[t]
\begin{center}
\epsscale{1.}
\plotone{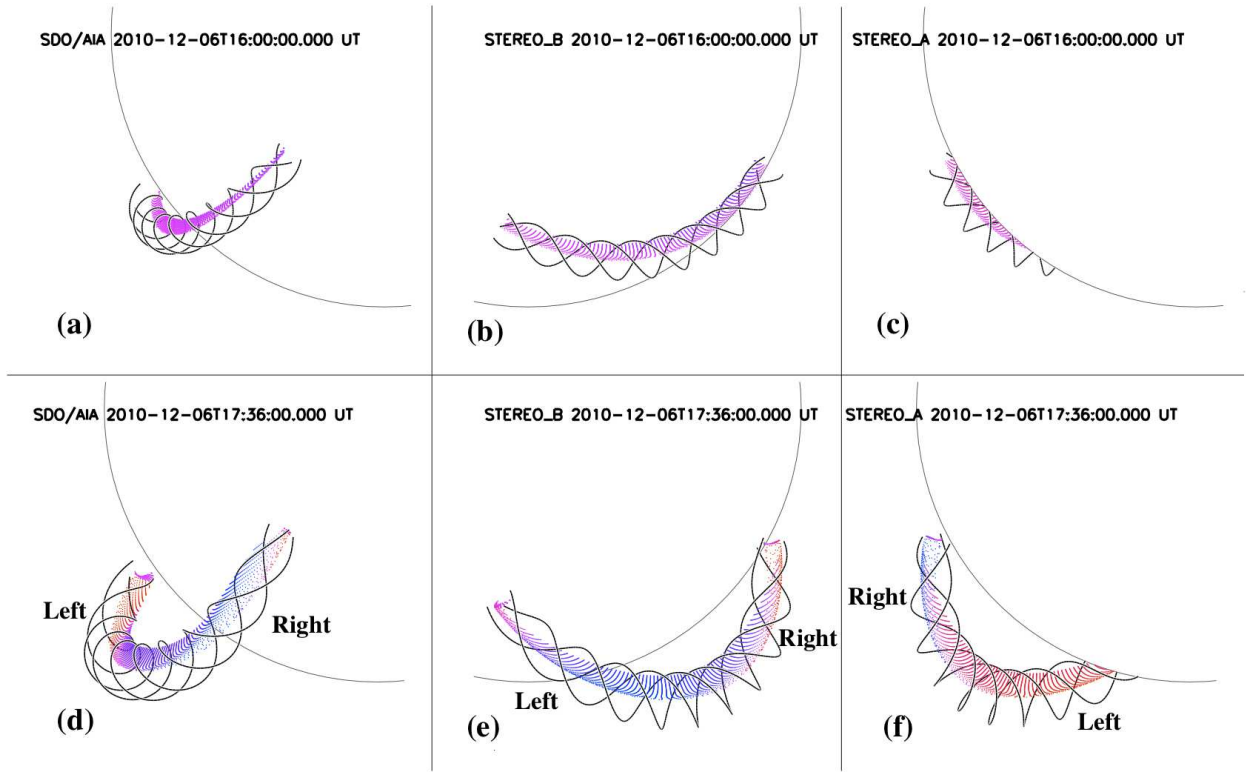}
\end{center}
\caption{Flux rope model for the 2010 December 6 prominence eruption
from the view of AIA (left column), STEREO$\_$B (middle column), and STEREO$\_$A (right column), respectively.
(a)--(c) Initial state at 16:00 UT:  the black curves indicate helical field lines at the outer edge of the flux rope, and the red line segments
simulate nearly vertical threads located in the lower half of the
helical flux rope. (d)--(f) Configuration at 17:36 UT when some of the
prominence blobs have been carried outward by the erupting flux rope,
while others are falling back down to the photosphere. The colors of
the blobs indicate the LOS velocity ($\pm 30$ $\rm km ~ s^{-1}$).
Animations of this figure (videos 5--7) with larger field of view are available online.}
\label{fig10}
\end{figure}

\begin{figure}[t]
\begin{center}
\epsscale{1.}
\plotone{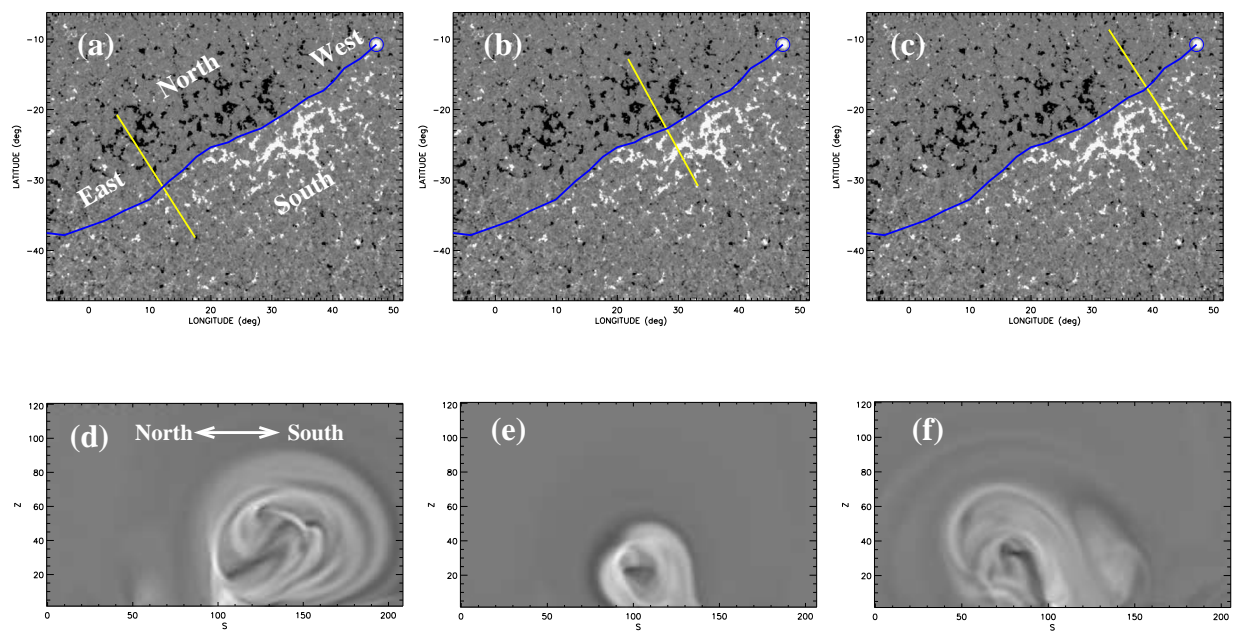}
\end{center}
\caption{Slices of $\alpha$ distribution from one NLFFF model (Model 1) in Paper~I. The background images in the top row are the maps of the radial component of the photospheric magnetic field observed by SDO/HMI. The zero point of the longitude corresponds to the central meridian on 2010 December 10 at 14:00 UT. The blue curve refers to the path where the flux rope is inserted. The bottom row shows vertical slices of $\alpha$ distribution along different parts of the flux rope (as indicated by the yellow line in the corresponding top-row image) from the NLFFF model.}
\label{fig11}
\end{figure}

\begin{figure}[t]
\begin{center}
\epsscale{0.8}
\plotone{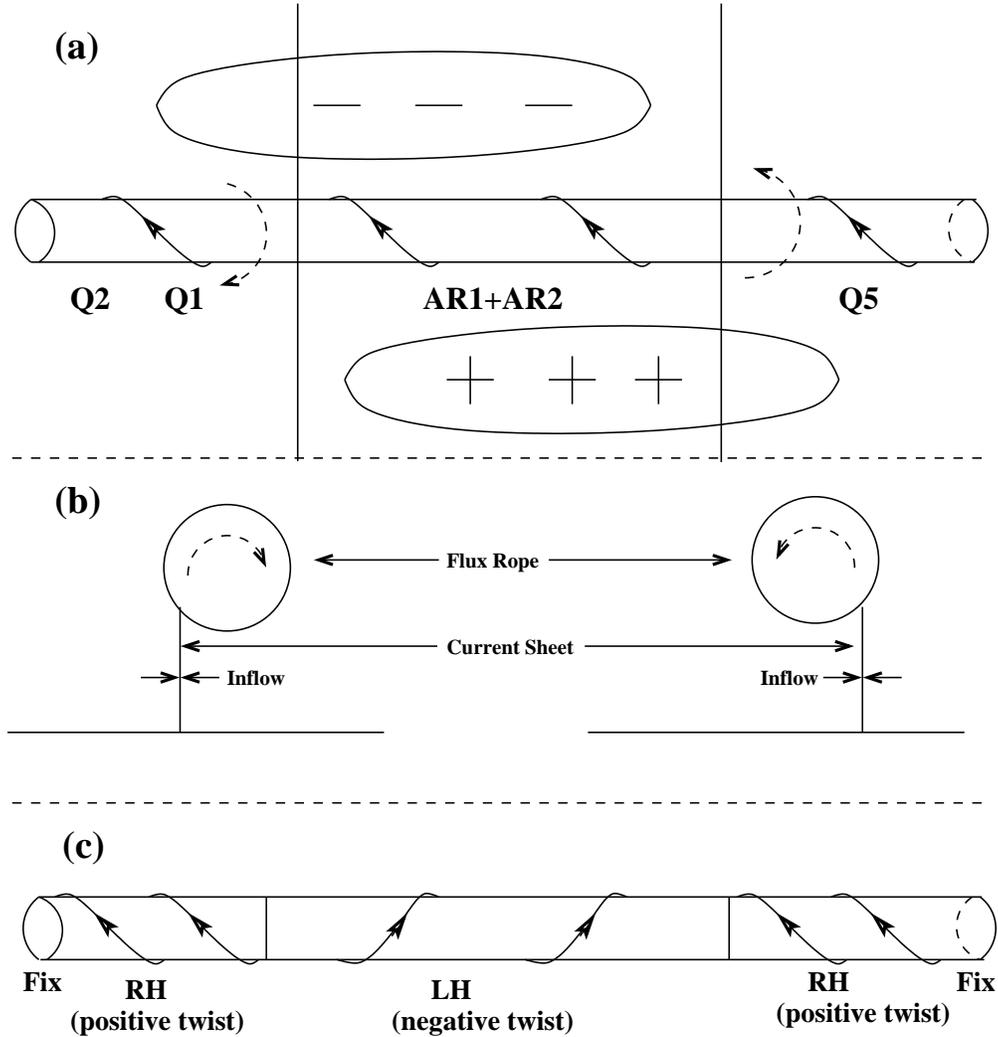}
\end{center}
\caption{Sketch of the formation of different twist along the flux rope at and near the active region.
(a) Before the eruption, the active region contains a weakly twisted flux rope
with left-pointing axial flux. (b) Cross sections of the asymmetric reconnection topology along different parts 
(as indicated by the two vertical lines in (a)) of the flux rope. (c) The asymmetric reconnection leads to rolling motions in
different directions as shown in the dashed arrows in (a) and (b), 
which lead to the formation of different twist along the flux rope.}
\label{fig12}
\end{figure}

\clearpage

\begin{table} 
\tabletypesize{\scriptsize}
\begin{center}
\caption{Activity timeline during the eruption. }
\begin{tabular}{cc}
\tableline\tableline
Time (UT) & Activity \\
\tableline  
2010-12-06 14:16 & Brightenings in active region \\   
2010-12-06 14:21 & Clear rise of AR1  \\   
2010-12-06 14:26 & Rolling motion of Q1 and Q2 begins  \\  
2010-12-06 14:46 & AR1 filament ejection in two directions  \\  
2010-12-06 15:30 & Clear Eruption of Q1 and Q2 begins  \\  
2010-12-06 16:00 & Rise of Q3 begins  \\ 
2010-12-06 16:00  & Bright curved features on the eastern quiet Sun begin to fade out  \\ 
2010-12-06 18:00  & AR2 eruption begins \\ 
2010-12-06 18:30  & Bright ribbons in active region appear \\ 
2010-12-06 19:30 & First post-eruption loops appear \\ 
2010-12-07 2:06  & Eruption of Q5 \\ 
\tableline 
\end{tabular}
\end{center}
\end{table}

 \begin{deluxetable}{llrrl}
\tablewidth{0pt}
\tablecaption{\label{table1}Model parameters.}
\tablehead{
\colhead{Parameter} & \colhead{Symbol} & \colhead{Model 1} &
\colhead{Model 2} & \colhead{Units} }
\startdata
Axis shape & $a_0$ & 0.28 & 0.28 &  \\
Exponent & $n$ & 3.57 & 3.57 &  \\
Initial height & $c_0$ & 0.05 & 0.16 & ${\rm R}_\odot$ \\
Flux rope radius & $R_0$ & 0.05 & 0.06 & ${\rm R}_\odot$ \\
Endpoint position & $d_0$ & 0.40 & 0.55 & ${\rm R}_\odot$ \\
 & $h_0$ & 0.981 & 0.906 & ${\rm R}_\odot$ \\
Outflow velocity & $v_{\rm max}$ & 5 & 70 & $\rm km ~ s^{-1}$ \\
Acceleration time & $\tau_{\rm acc}$ & 300 & 3600 & s \\
Writhing velocity & $v_{\rm writhe}$ & 0 & $-10$ & $\rm km ~ s^{-1}$ \\
Writhing timescale & $\tau_{\rm writhe}$ & -- & 3600 & s \\
Friction & $\beta$ & 0.01 & 0.001 & $\rm s^{-1}$ \\
Helical windings: & & & & \\
~~~ Initial number & $N_0$ & $+1$ & $+3$ &  \\
~~~ Rate of increase & $N_1$ & $0.4$ & $0.2$ & $\rm hr^{-1}$ \\
~~~ Zero point & $\xi_0$ & $-1$ & $-1$ &  \\
Source latitude & -- & $-35$ & $-35$ & deg \\
Source longitude & -- & $-40$ & $-55$ & deg \\
Tilt w.r.t.~latitude circle & -- & $-35$ & $-10$ & deg \\
Tilt w.r.t.~radial (NS) & -- & 0 & 20 & deg \\
SDO latitude & -- & $+0.192$ & $+0.192$ & deg \\
STEREO-A: & & & & \\
~~~ Longitude & -- & $+85.993$ & $+85.993$ & deg \\
~~~ Latitude & -- & $-7.308$ & $-7.308$ & deg \\
STEREO-B: & & & & \\
~~~ Longitude & -- & $-86.566$ & $-86.566$ & deg \\
~~~ Latitude & -- & $+7.305$ & $+7.305$ & deg 
\enddata
\end{deluxetable}


\clearpage

                                                            
\begin{thebibliography}{}   

\bibitem[Alexander et al.(2006)]{2006ApJ...653..719A} Alexander, D., Liu, 
R., \& Gilbert, H.~R.\ 2006, \apj, 653, 719 

\bibitem[Ali et al.(2007)]{2007SoPh..240...89A} Ali, S.~S., Uddin, W., 
Chandra, R., Mary, D.~L., \& Vr{\v s}nak, B.\ 2007, \solphys, 240, 89 

\bibitem[Amari et al.(2003)]{2003ApJ...585.1073A} Amari, T., Luciani, 
J.~F., Aly, J.~J., Mikic, Z., \& Linker, J.\ 2003, \apj, 585, 1073 

\bibitem[Antiochos et al.(1999)]{1999ApJ...510..485A} Antiochos, S.~K., 
DeVore, C.~R., \& Klimchuk, J.~A.\ 1999, \apj, 510, 485 

\bibitem[Antolin  \& Rouppe van der Voort(2012)]{2012ApJ...745..152A} Antolin, P., \& Rouppe van der Voort, L.\ 2012, \apj, 745, 152 

\bibitem[Bemporad et 
al.(2011)]{2011A&A...531A.147B} Bemporad, A., Mierla, M., \& Tripathi, D.\ 2011, \aap, 531, A147 


\bibitem[Berger  \& Field(1984)]{1984JFM...147..133B} Berger, M.~A., \& Field, G.~B.\ 1984, Journal of Fluid Mechanics, 147, 133 

\bibitem[Burlaga(1991)]{1991pihp.book....1B} Burlaga, L.~F.~E.\ 1991, 
Physics of the Inner Heliosphere II, 1 


\bibitem[Chen et al.(1997)]{1997ApJ...490L.191C} Chen, J., Howard, R.~A., 
Brueckner, G.~E., et al.\ 1997, \apjl, 490, L191 

\bibitem[Cohen et al.(2010)]{2010JGRA..11510104C} Cohen, O., Attrill, 
G.~D.~R., Schwadron, N.~A., et al.\ 2010, Journal of Geophysical Research 
(Space Physics), 115, 10104 

\bibitem[Fan 
\& Gibson(2007)]{2007ApJ...668.1232F} Fan, Y., \& Gibson, S.~E.\ 2007, \apj, 668, 1232 


\bibitem[Forbes(2000)]{2000JGR...10523153F} Forbes, T.~G.\ 2000, \jgr, 105, 23153

\bibitem[Forbes 
\& Isenberg(1991)]{1991ApJ...373..294F} Forbes, T.~G., \& Isenberg, P.~A.\ 1991, \apj, 373, 294 

\bibitem[Gibson 
\& Low(1998)]{1998ApJ...493..460G} Gibson, S.~E., \& Low, B.~C.\ 1998, \apj, 493, 460 

\bibitem[Gilbert et al.(2007)]{2007SoPh..245..287G} Gilbert, H.~R., 
Alexander, D., \& Liu, R.\ 2007, \solphys, 245, 287 

\bibitem[Green et al.(2007)]{2007SoPh..246..365G} Green, L.~M., Kliem, B., 
T{\"o}r{\"o}k, T., van Driel-Gesztelyi, L., 
\& Attrill, G.~D.~R.\ 2007, \solphys, 246, 365 

\bibitem[Golub et al.(2007)]{2007SoPh..243...63G} Golub, L., et al.\ 2007, 
\solphys, 243, 63 

\bibitem[Harrison(1996)]{1996SoPh..166..441H} Harrison, R.~A.\ 1996, \solphys, 166, 441

\bibitem[Howard et al.(2008)]{2008SSRv..136...67H} Howard, R.~A., et al.\ 
2008, \ssr, 136, 67 

\bibitem[Ji et al.(2003)]{2003ApJ...595L.135J} Ji, H., Wang, H., Schmahl, 
E.~J., Moon, Y.-J., \& Jiang, Y.\ 2003, \apjl, 595, L135 

\bibitem[Joshi 
\& Srivastava(2011)]{2011ApJ...730..104J} Joshi, A.~D., \& Srivastava, N.\ 2011, \apj, 730, 104 

\bibitem[Kano et al.(2008)]{2008SoPh..249..263K} Kano, R., et al.\ 2008, 
\solphys, 249, 263 

\bibitem[Kliem \&
T{\"o}r{\"o}k(2006)]{2006PhRvL..96y5002K} Kliem, B., T{\"o}r{\"o}k, T.\ 2006, Physical Review Letters, 96, 255002


\bibitem[Kliem et al.(2012)]{2012SoPh..tmp...91K} Kliem, B., T{\"o}r{\"o}k, 
T., \& Thompson, W.~T.\ 2012, \solphys, 91 

\bibitem[Koleva et 
al.(2012)]{2012A&A...540A.127K} Koleva, K., Madjarska, M.~S., Duchlev, P., et al.\ 2012, \aap, 540, A127 
  
\bibitem[Kosugi et al.(2007)]{2007SoPh..243....3K} Kosugi, T., et al.\ 
2007, \solphys, 243, 3 

\bibitem[Krall et al.(2000)]{2000ApJ...539..964K} Krall, J., Chen, J., 
\& Santoro, R.\ 2000, \apj, 539, 964 

\bibitem[Kurokawa et al.(1987)]{1987SoPh..108..251K} Kurokawa, H., Hanaoka, 
Y., Shibata, K., \& Uchida, Y.\ 1987, \solphys, 108, 251 


\bibitem[Lemen et al.(2012)]{2012SoPh..275...17L} Lemen, J.~R., Title, 
A.~M., Akin, D.~J., et al.\ 2012, \solphys, 275, 17 


\bibitem[Liu \& Alexander(2009)]{2009ApJ...697..999L} Liu, R., \& Alexander, D.\ 2009, \apj, 697, 999 

\bibitem[Liu et al.(2007)]{2007ApJ...661.1260L} Liu, R., Alexander, D., 
\& Gilbert, H.~R.\ 2007, \apj, 661, 1260 

\bibitem[Liu et al.(2012)]{2012ApJ...756...59L} Liu, R., Kliem, B., 
T{\"o}r{\"o}k, T., et al.\ 2012, \apj, 756, 59 

\bibitem[Manchester(2003)]{2003JGRA..108.1162M} Manchester, W.\ 2003, 
Journal of Geophysical Research (Space Physics), 108, 1162

\bibitem[Martin(2003)]{2003AdSpR..32.1883M} Martin, S.~F.\ 2003, Advances 
in Space Research, 32, 1883 

\bibitem[Mikic 
\& Linker(1994)]{1994ApJ...430..898M} Mikic, Z., \& Linker, J.~A.\ 1994, \apj, 430, 898 

\bibitem[Mikic et al.(1988)]{1988ApJ...328..830M} Mikic, Z., Barnes, D.~C., 
\& Schnack, D.~D.\ 1988, \apj, 328, 830 

\bibitem[Muglach et al.(2009)]{2009ApJ...703..976M} Muglach, K., Wang, 
Y.-M., \& Kliem, B.\ 2009, \apj, 703, 976 

\bibitem[Murphy et al.(2012)]{2012ApJ...751...56M} Murphy, N.~A., Miralles, 
M.~P., Pope, C.~L., et al.\ 2012, \apj, 751, 56 

\bibitem[Panasenco 
\& Martin(2008)]{2008ASPC..383..243P} Panasenco, O., \& Martin, S.~F.\ 2008, Subsurface and Atmospheric Influences on Solar Activity, 383, 243 

\bibitem[Panasenco et al.(2011)]{2011JASTP..73.1129P} Panasenco, O., 
Martin, S., Joshi, A.~D., 
\& Srivastava, N.\ 2011, Journal of Atmospheric and Solar-Terrestrial Physics, 73, 1129 

\bibitem[Reeves et al.(2008)]{2008ApJ...675..868R} Reeves, K.~K., Seaton, 
D.~B., \& Forbes, T.~G.\ 2008, \apj, 675, 868 


\bibitem[Rompolt(1975)]{1975STIN...7614007R} Rompolt, B.\ 1975, NASA 
STI/Recon Technical Report N, 76, 14007 

\bibitem[Roussev et al.(2003)]{2003ApJ...588L..45R} Roussev, I.~I., Forbes, 
T.~G., Gombosi, T.~I., et al.\ 2003, \apjl, 588, L45 

\bibitem[Rust(2001)]{2001JGR...10625075R} Rust, D.~M.\ 2001, \jgr, 106, 
25075 

\bibitem[Rust(2003)]{2003AdSpR..32.1895R} Rust, D.~M.\ 2003, Advances in 
Space Research, 32, 1895 

\bibitem[Rust \& Kumar(1994)]{1994SoPh..155...69R} Rust, D.~M., \& Kumar, A.\ 1994, \solphys, 155, 69 

\bibitem[Schou et al.(2012)]{2012SoPh..275..229S} Schou, J., Scherrer, 
P.~H., Bush, R.~I., et al.\ 2012, \solphys, 275, 229 

\bibitem[Srivastava et al.(1991)]{1991SoPh..133..339S} Srivastava, N., 
Ambastha, A., \& Bhatnagar, A.\ 1991, \solphys, 133, 339 

\bibitem[Sterling \& Moore(2005)]{2005ApJ...630.1148S} Sterling, A.~C., \& Moore, R.~L.\ 2005, \apj, 630, 1148 

\bibitem[Su et al.(2012)]{2012ApJ...756L..41S} Su, Y., Wang, T., Veronig, 
A., Temmer, M., \& Gan, W.\ 2012, \apjl, 756, L41   

\bibitem[Su \& van Ballegooijen(2012)]{2012ApJ...757..168S} Su, Y., \& van Ballegooijen, A.\ 2012, \apj, 757, 168, Paper~I. 

\bibitem[Su et al.(2010)]{2010ApJ...721..901S} Su, Y., van Ballegooijen, A., \& Golub, L.\ 2010, \apj, 721, 901 

\bibitem[Su et al.(2012)]{2012ASPC..454..113S} Su, Y., van Ballegooijen, A., \& Golub, L.\ 2012, Hinode-3: The 3rd Hinode Science Meeting, 454, 113 

\bibitem[Sun et al.(2012)]{2012ApJ...757..149S} Sun, X., Hoeksema, J.~T., 
Liu, Y., Chen, Q., \& Hayashi, K.\ 2012, \apj, 757, 149 

\bibitem[T{\"o}r{\"o}k et 
al.(2010)]{2010A&A...516A..49T} T{\"o}r{\"o}k, T., Berger, M.~A., \& Kliem, B.\ 2010, \aap, 516, A49 

\bibitem[T{\"o}r{\"o}k et 
al.(2004)]{2004A&A...413L..27T} T{\"o}r{\"o}k, T., Kliem, B., \& Titov, V.~S.\ 2004, \aap, 413, L27 

\bibitem[T{\"o}r{\"o}k et al.(2008)]{2008ESPM...12.3.54T} T{\"o}r{\"o}k, 
T., Berger, M.~A., Kliem, B., et al.\ 2008, ''12th European Solar Physics 
Meeting, Freiburg, Germany, held September, 8-12, 2008.~Online at 
``http://espm.kis.uni-freiburg.de/, p.3.54'', 12, 3 

\bibitem[Vrsnak(1990)]{1990SoPh..127..129V} Vrsnak, B.\ 1990, \solphys, 
127, 129 

\bibitem[Vrsnak et al.(1991)]{1991SoPh..136..151V} Vrsnak, B., Ruzdjak, V., 
\& Rompolt, B.\ 1991, \solphys, 136, 151 


\bibitem[Vrsnak et al.(1988)]{1988SoPh..116...45V} Vrsnak, B., Ruzdjak, V., 
Brajsa, R., \& Dzubur, A.\ 1988, \solphys, 116, 45 

\bibitem[Vrsnak et al.(1993)]{1993SoPh..146..147V} Vrsnak, B., Ruzdjak, V., 
Rompolt, B., Rosa, D., \& Zlobec, P.\ 1993, \solphys, 146, 147 

\bibitem[Wang 
\& Shi(1993)]{1993ASPC...46..397W} Wang, J., \& Shi, Z.\ 1993, IAU Colloq.~141: The Magnetic and Velocity Fields of Solar Active Regions, 46, 397 

\bibitem[Williams et al.(2005)]{2005ApJ...628L.163W} Williams, D.~R., 
T{\"o}r{\"o}k, T., D{\'e}moulin, P., van Driel-Gesztelyi, L., 
\& Kliem, B.\ 2005, \apjl, 628, L163 

\bibitem[Thompson(2011)]{2011JASTP..73.1138T} Thompson, W.~T.\ 2011, 
Journal of Atmospheric and Solar-Terrestrial Physics, 73, 1138 

\bibitem[Thompson(2012)]{Thompson2012} Thompson, W.~T.\ 2012, \solphys, submitted
 
\bibitem[Wu et al.(1997)]{1997SoPh..170..265W} Wu, S.~T., Guo, W.~P., 
\& Dryer, M.\ 1997, \solphys, 170, 265 

\bibitem[Wuelser et al.(2004)]{2004SPIE.5171..111W} Wuelser, J.-P., et al.\ 
2004, \procspie, 5171, 111 

\bibitem[Zhou et al.(2006)]{2006ApJ...651.1238Z} Zhou, G.~P., Wang, J.~X., 
Zhang, J., et al.\ 2006, \apj, 651, 1238 


\end{thebibliography}
\end{document}